\documentclass[linenumbers, aps, prb, twocolumn, epsfig, amssymb,tighten]{revtex4}
\usepackage{epstopdf,xcolor,graphicx,hyperref,physics,amsmath,comment}
\usepackage{color}
\usepackage{verbatimbox}

\usepackage[caption=false]{subfig}
\hypersetup{colorlinks=true,
    linkcolor=blue,
    citecolor=blue,
    filecolor=magenta,      
    urlcolor=blue,}
\renewcommand{\r}{\vb{r}}

\newcommand{\lD}{\lambda_D}
\renewcommand{\k}{\vb{k}}
\newcommand{\ck}{{c}_{\k}}
\newcommand{\ckd}{\ck^{\dagger}}
\newcommand{\half}{\frac{1}{2}}
\newcommand{\flathalf}{\flatfrac{1}{2}}
\renewcommand{\d}{\mathbf{d}}
\newcommand{\bsigma}{\boldsymbol{\sigma}}

\newcommand{\m}{\vb{{m}}}
\newcommand{\mz}{{m}_z}

\newcommand{\Z}{\mathcal{Z}}

\newcommand{\FM}{F_M}

\newcommand{\Bs}{B_{s}}

\newcommand{\BesselK}[2]{\mathcal{K}_{#1}\qty(#2)}

\renewcommand{\l}{\lambda}
\newcommand{\xmu}{x_{\mu}}

\newcommand{\zh}{\vb{\hat{z}}}

\newcommand{\taus}{\tau^*}
\newcommand{\Ks}{K^*}
\newcommand{\lO}{\lambda_0}
\newcommand{\ci}{c_{i}}
\newcommand{\cid}{\ci^\dagger}

\newcommand{\xh}{\hat{\mathbf{x}}}
\begin{document}
\title{Tunable skyrmion-skyrmion binding on the surface of a topological insulator}
\author{Kunal L.~Tiwari, J.~Lavoie, T.~Pereg-Barnea, W.~A.~Coish}
\affiliation{Department of Physics, McGill University, Montr\'eal, Qu\'ebec,
Canada H3A 2T8}

\date{\today}
\begin{abstract}
We show that skyrmions on the surface of a magnetic topological insulator may experience an attractive interaction that leads to the formation of a skyrmion-skyrmion bound state. This is in contrast to the case of skyrmions in a conventional chiral ferromagnet, for which the intrinsic interaction is repulsive.  The origin of skyrmion binding in our model is the molecular hybridization of topologically protected electronic orbitals associated with each skyrmion. Attraction between the skyrmions can therefore be controlled by tuning a chemical potential that populates/depopulates the lowest-energy molecular orbital. We find that the skyrmion-skyrmion bound state can be made stable, unstable, or metastable depending on the chemical potential, magnetic field, and easy-axis anisotropy of the underlying ferromagnet, resulting in a rich phase diagram.  Finally, we discuss the possibility to realize this effect in a recently synthesized Cr doped ${\qty(\mathrm{Bi}_{2-y}\mathrm{Sb}_{y})}_{2}\mathrm{Te}_3$ heterostructure.
\end{abstract}
\maketitle

\section{Introduction}

The surface of a strong three-dimensional topological insulator hosts a set of two-dimensional surface states protected by topology, provided that the surface does not break the protecting symmetries.\cite{2010_Hasan, 2016_Bansil, 2016_Chiu} Within the bulk gap, these states are characterized by a chiral Dirac cone dispersion. When time-reversal symmetry is broken, these surface states are no longer protected and may be gapped. In a magnetic topological insulator (MTI), magnetic moments result in broken time-reversal symmetry.  The magnetic moments may couple to each other through direct or indirect [Ruderman-Kittel-Kasuya-Yosida (RKKY)] exchange and form an ordered state.\cite{2009_Liu, 2011_Zhu} These moments may also couple to the electronic subsystem through a Zeeman-like term proportional to the local magnetization. In the case of uniform out-of-plane ferromagnetic order, the magnetization gives the surface Dirac electrons a finite mass, resulting in a gap in the surface-state spectrum.
 
The gapped surface states can be effectively described by a massive Dirac model. Therefore, any sign change of the Dirac mass leads to localized Jackiw-Rebbi modes \cite{2019_Tokura,1976_Jackiw,1984_Jackiw}.  Consequently, when the massive Dirac electrons are coupled to ferromagnetic moments there will be a set of protected one-dimensional edge states associated with the boundary between magnetic domains with opposite magnetization. Such edge states are responsible for the anomalous quantum Hall effect,\cite{2013_Chang} and are thought to give rise to butterfly hysteresis in the magnetotransport properties of magnetic topological insulators.\cite{2015_Nakajima, 2017_Tiwari} Magnetic skyrmions, topological defects in the magnetization, give rise to similar topologically protected electronic states at the skyrmion perimeter.\cite{2015_Hurst}

\begin{figure}[!]
    \includegraphics{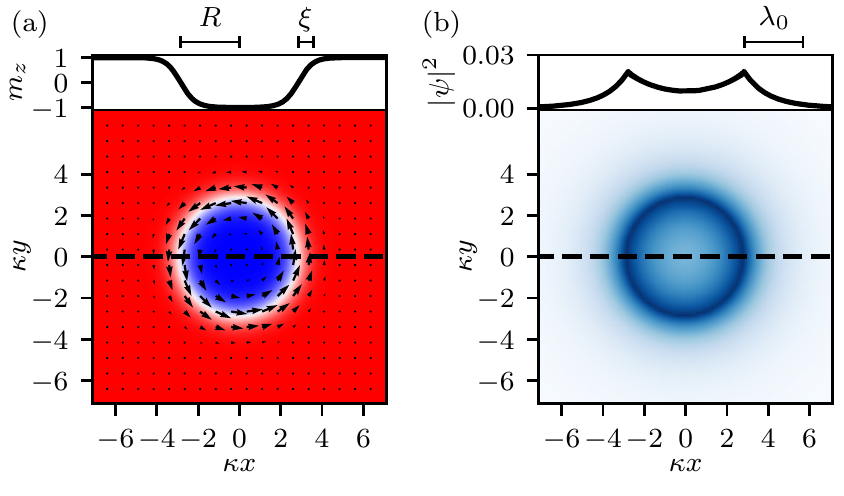}
    \caption{(a) The magnetic texture, $\m = {\qty(m_x, m_y, m_z)}^{T}$, of an isolated skyrmion, and (b), the probability density, $\abs{\psi\qty(\r)}^2$, of an electronic state bound to the skyrmion. In (a), the color scale represents $\mz$, ranging from $\mz=-1$ (blue) to $\mz =1 $ (red). Arrows indicate the in-plane component of the magnetization, ${\qty(m_x, m_y)}^{T}$. The upper inset shows $\mz$ along the dashed line. The skyrmion radius $R$, defined through Eq.~\eqref{eq_Rdef}, and healing length $\xi$, defined through Eq.~\eqref{eq_xidef}, are the radius and width of the white ring, respectively. The magnetization plotted in (a) was determined numerically by solving Eq.~\eqref{eq_FMfdv} using the procedure described in Appendix \ref{ap_2sknumerics} with boundary conditions corresponding to a single skyrmion. In (b) we plot the probability density of the lowest-energy in-gap electronic orbital for $\lO=R$ [with $\lO$ defined in Eq.~\eqref{eq_l0defn}]. The wavefunction, $\psi$, was determined using the procedure described in Appendix \ref{ap_electronic} with the magnetization plotted in (a).  Here, $\psi$ corresponds to the $j=\flathalf$ state of the continuum model [see Eqs.~\eqref{eq_totalwavefn},~\eqref{eq_radial_wave_eqn}]. We find numerically that this state has energy $\flatfrac{E}{\Delta}=-0.614$. The probability density, $\abs{\psi}^2$, along the slice indicated by the dashed line is plotted in the top panel.\label{fig_1sk_cmaps}}
\end{figure}

Magnetic skyrmions are localized topologically stable configurations of the magnetization for chiral ferromagnets.\cite{2013_Nagaosa, 2006_Rossler} We consider skyrmions in a planar magnetic system having an easy-axis anisotropy and in the presence of a
finite applied magnetic field perpendicular to the surface (along $\zh$), both of which tend to stabilize the skyrmion phase.\cite{2006_Rossler} Far
from a skyrmion, the magnetization, $\m$, is uniformly aligned with the applied
magnetic field, $\m \parallel\zh$. At the center of a skyrmion, the
magnetization is anti-aligned with the applied magnetic field, $\m
\parallel-\zh$. Across the perimeter of a skyrmion, moving radially outward, $m_z$ smoothly interpolates between these boundary conditions as the magnetization vector tilts in-plane, with the in-plane component of magnetization perpendicular to the to the radial direction $\hat{r}$ (for Bloch-type skyrmions) [Fig.~\ref{fig_1sk_cmaps}\!(a)]. The magnetization vector $\m\qty(\r)$ at position $\r$ wraps the unit sphere as $\r$ traverses the entire plane. This wrapping is a manifestation of the skyrmion's topological nature; the magnetization associated with a skyrmion cannot
be smoothly deformed to a uniform ferromagnetic state.  This fact, coupled with
a ferromagnetic exchange interaction that prohibits sharp changes in the
magnetization, makes skyrmions exceptionally stable. This stability is present
even if the skyrmion is higher in energy than the uniform ferromagnetic phase.  In Ref.~[\onlinecite{2015_Hurst}], Hurst \emph{et al.}~consider a topological insulator surface coupled to a planar magnet that hosts a skyrmion. They find a discrete set of protected orbitals bound to the skyrmion [see, e.g., Fig.~\ref{fig_1sk_cmaps}\!(b)].  Similar electronic states have also been investigated in related systems.\cite{2013_Ferreira,2015_Uchoa} 

In this paper, we consider the interaction between a pair of skyrmions on the surface of a MTI. In the absence of the Dirac surface states, a pair of skyrmions in a chiral ferromagnet experiences a mutually repulsive interaction at all distances.\cite{2013_Lin} This repulsive interaction decays exponentially with the inter-skyrmion separation over the magnetic healing length, $\xi$. The healing length is controlled by the applied magnetic field and easy-axis anisotropy. Once the Dirac electronic system is coupled to the magnetic system, skyrmion-bound electronic states form.  Their wave functions overlap and hybridize when two skyrmions are close.  If an electron occupies the lowest-lying hybridized electronic state, there will be an attractive contribution to the skyrmion-skyrmion interaction from the associated molecular binding energy. This attractive interaction also decays exponentially with separation and is governed independently by the skyrmion-bound orbital decay length. Occupation of the lowest-energy electronic orbital, and hence the attractive interaction, can be controlled by coupling the electronic system to a reservoir and tuning its chemical potential. Frustrated magnetic systems, with competing antiferromagnetic and ferromagnetic couplings, may show an oscillating repulsive/attractive interaction between skyrmions as a function of distance, even in the absence of an electronic system.\cite{2015_Leonov,2017_Kharkov}  In this work, in contrast, our focus is on chiral ferromagnetic systems, where the magnetic interaction is purely repulsive.

A central result of this work is a zero-temperature skyrmion-skyrmion binding phase diagram, displaying the stability of bound skyrmion pairs as a function of the electrons' chemical potential and magnetic healing length.  The healing length may be controlled through the applied magnetic field and the easy-axis anisotropy. We also extend these results to low, but finite, temperature and establish concrete conditions to realize this effect experimentally.  Understanding and controlling magnetic skyrmion binding through the electronic system may be important for both classical skyrmionic devices\cite{2017_Fert} and for possible qubit implementations.\cite{2013_Ferreira} 

The remainder of this paper is structured as follows. In
Section~\ref{sec_msystem}, we review the theory of individual magnetic
skyrmions. Evaluating the magnetic free energy of a system containing two skyrmions as a
function of skyrmion-skyrmion separation reveals a short-range repulsive interaction. In Section~\ref{sec_esystem}, we
consider the electronic subsystem. Topologically protected
electron orbitals bound to single skyrmions hybridize to form molecular orbitals
in a two-skyrmion system. We determine the grand potential of this
system in contact with an electronic reservoir as a function of
skyrmion-skyrmion separation and conclude that the electronic system gives rise to an attractive interaction. In Section~\ref{sec_phasediagram}, we analyze the
total skyrmion-skyrmion interaction as a function of tunable parameters, finding a phase diagram for skyrmion-skyrmion binding at zero temperature, and then
addressing the case of low, but finite, temperature. Finally, in
Section~\ref{sec_considerations}, we discuss the possibility to realize this effect in 
$\textrm{Cr}_{x}{(\textrm{Bi}_{1-y}\textrm{Sb}_{y})}_{2-x}\textrm{Te}_{3} /
{(\textrm{Bi}_{1-y}\textrm{Sb}_{y})}_{2}\textrm{Te}_{3}$ and justify the approximations we have made in the context of this material.

\section{Short-range magnetic repulsion}
\label{sec_msystem}
In this section, we consider the magnetic subsystem in isolation to determine
the range and character of the intrinsic repulsive skyrmion-skyrmion
interaction. Starting from a standard free energy for a
planar chiral magnet, with the addition of an easy-axis anisotropy, we determine
the single-skyrmion magnetization. The skyrmion is then characterized by two
length scales: the radius, $R$, and the healing length, $\xi$ [see
Fig.~\ref{fig_1sk_cmaps}\!(a)]. Next, we numerically determine the magnetic free
energy of a system of two skyrmions as a function of inter-skyrmion
separation. This analysis reveals that the magnetic repulsive
interaction decays exponentially with inter-skyrmion separation over the healing
length, $\xi$. If $\xi$ is sufficiently short, the repulsion will be
overcome by a longer-range attractive interaction due to the electronic
subsystem.

The standard free energy for a two-dimensional chiral magnet stabilizing skyrmions includes the 
ferromagnetic exchange interaction, the Dzyaloshinskii-Moriya interaction (DMI),
and a perpendicular applied magnetic field.\cite{2013_Nagaosa} We consider this
magnetic free energy with an additional term accounting for an easy-axis
anisotropy, which may stabilize skyrmions in the absence of an applied magnetic field.
The free energy density, $f$, and total magnetic free energy, $F_M$, are
\begin{align}
    f &= \frac{J}{2}{\qty(\grad\m)}^{2} + D\m\cdot\qty(\grad\times\m)  -B \mz -
    K\mz^{2},\label{eq_magf_bare}\\
    \FM &= \int \dd[2]{\r}f\label{eq_FMtotal}.
\end{align}
Here, $J>0$ is proportional to the exchange constant, $D$ arises from the
DMI,\cite{1958_Dzyaloshinsky,1960_Moriya} $B$ is proportional to the
out-of-plane applied magnetic field, and $K$ is the easy-axis anisotropy.  The
magnetic system has a natural inverse length scale $\kappa$, and a natural
energy scale $B_s$:
\begin{align}
    \kappa &= \frac{D}{J}, \\
    \Bs &= \frac{D^2}{J}.
\end{align}

We neglect fluctuations in $\abs{\m}$, which may be penalized by terms quadratic
and quartic in $\abs{\m}$ (not explicitly included here).  Without loss of
generality, we set $\abs{\m}=1$. For $D=0$ and $B>0$, the ground state of the
magnetic system is uniform, with $\m = \zh$ throughout
the plane.  For finite $D$ and sufficiently large $B$ and/or $K$, skyrmions may
exist as locally stable features in the magnetization.\cite{2006_Rossler}
The presence of a skyrmion in the ferromagnetic background may
either increase or decrease the magnetic free energy depending on $B$ and $K$. A
discussion of the phase diagram of chiral magnets governed by
Eq.~\eqref{eq_magf_bare} is given in Refs.~[\onlinecite{2016_Utkan,
2014_Banerjee}]. 

\begin{figure}
    \includegraphics{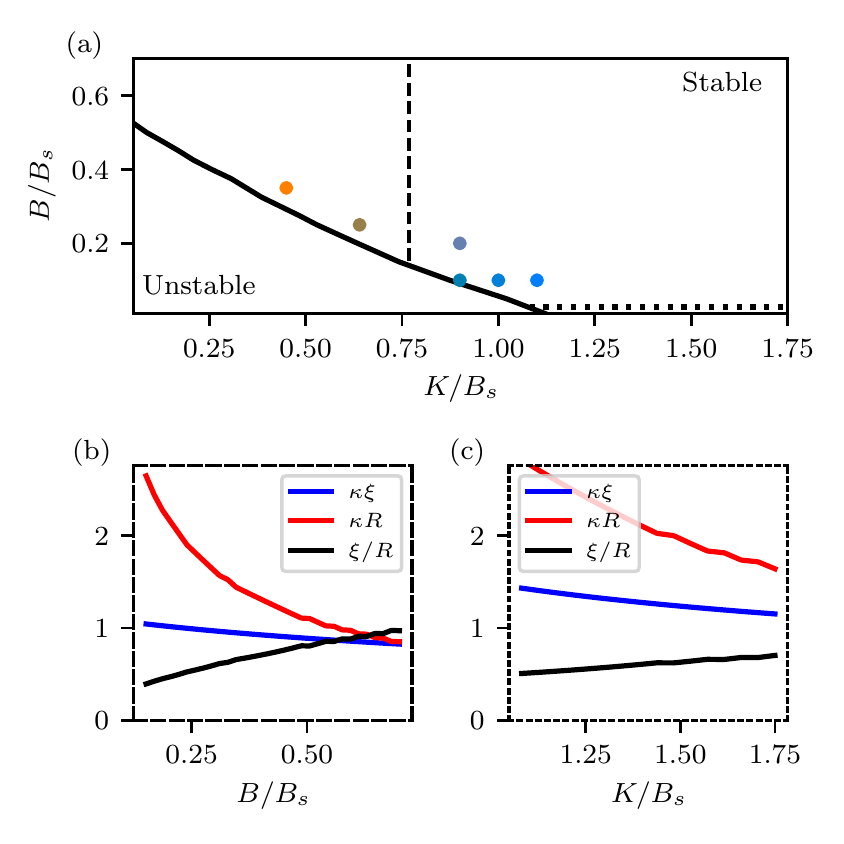}
    \caption{Skyrmion stability phase diagram in the $K,B$ plane. (a) for small
    $K$ and $B$, the magnetic system is unstable toward a spin spiral
    phase. Above this phase boundary (black line), isolated skyrmions
    minimize the magnetic free energy. Colored dots indicate the values of $B,K$ used to perform numerical calculations in the stable phase for Figs.~\ref{fig_FMofX},\ref{fig_hybridization} below (see Table~\ref{table_nparams}). The phase boundary was determined using the procedure described in Appendix \ref{ap_stability}. In (b) and (c), we plot the healing
    length, $\xi$, (blue), the skyrmion radius, $R$, (red), and the ratio $\xi/R$ (black) as a function
    of applied magnetic field (b) and anisotropy (c)
    for the slices indicated in (a) with a dashed and dotted line, respectively. The ratio $\xi/R$,
    which must be small for a skyrmion bound state to form, is minimized for
    small applied field and anisotropy approaching the phase boundary  
    }\label{fig_magnetic_parameters}
\end{figure}

For sufficiently large $B$ and/or $K$, a magnetic skyrmion minimizes the
magnetic free energy, i.e., it solves
\begin{align}
    \fdv{\FM}{\m}=&0\label{eq_FMfdv},
\end{align} 
with the boundary conditions $\m\qty(r=0) = -\zh$ and
$\m{\qty(r\rightarrow\infty)} = \zh$ for positive $B$. These boundary conditions
ensure that the skyrmion is centered at $r=0$, and that the magnetization far
from the skyrmion approaches the value it would have in the uniform
ferromagnetic phase. The magnetization profile of the skyrmion may be
parameterized as
\begin{align}
    \m\qty(\r)=
    \mqty(
    \sin\qty[\Theta\qty(r)]\cos\qty(W\phi+\phi_0)\\
    \sin\qty[\Theta\qty(r)]\sin\qty(W\phi+\phi_0)\\ 
    \cos\qty[\Theta\qty(r)]),
\end{align}
where $W$ is the winding number of the skyrmion, and $\phi_0$ determines its chirality. For $D>0$, the specific form of the DMI considered in Eq.~\eqref{eq_magf_bare} stabilizes spiral-like (Bloch) skyrmions, with a definite chirality, $\phi_0=+\flatfrac{\pi}{2}$,\footnote{More generally, the DMI for a two-dimensional thin film can be written as $f_{DM}=D_1\mathbf{m}\cdot\left(\nabla\times\mathbf{m}\right)+D_2\left[\left(\mathbf{m}\cdot\nabla\right)m_z-m_z\left(\nabla\cdot\mathbf{m}\right)\right]$.  For concrete calculations in this manuscript, we have chosen $D_1=D>0$, $D_2=0$, leading to stable Bloch-type skyrmions with $\phi_0=+\pi/2$.  In the more general case ($D_1\ne 0,\,D_2\ne 0$), some other (fixed) value of $\phi_0$ will be stabilized. For any value of $\phi_0$, the qualitative arguments leading to magnetic repulsion hold [see the discussion following Eq.~\eqref{eq_Rdef}, below].  Because the electronic states are topological in origin, we expect their presence to be robust to changes in the in-plane magnetization, so the electronic interaction is also robust to changes in the type of skyrmion (value of $\phi_0$), although the wavefunctions may be subject to a more complicated wave equation, Eq.~\eqref{eq_radial_wave_eqn}.  We therefore expect our qualitative analysis of skyrmion binding to hold for skyrmions of N\'eel-type ($\phi_0=0$), Bloch-type ($\phi_0=\pi/2$), and intermediate type [$\phi_0 \in (0,\pi/2)$].} and a definite winding number, $W=+1$
(see, e.g., Ref.~\onlinecite{2017_Han}).\footnote{This is in contrast with the case of a centrosymmetric frustrated magnetic system, not considered here, where skyrmions of either winding number, $W=\pm 1$, may be stable.\cite{2016_Lin}} An individual skyrmion satisfies the equation
\begin{align}
    \fdv{\FM}{\Theta} = 0, \label{eq_1skODE}
\end{align}
with the boundary condition $\Theta\,\qty(r=0)=\pi$,
$\Theta\qty(r\rightarrow\infty)=0$.
Away from the skyrmion core, for $\Theta \ll 1$ and $\kappa r \gg 1$, Eq.~\eqref{eq_1skODE} is solved
asymptotically by 
\begin{align}
    \Theta\qty(r) \propto  \BesselK{1}{\frac{r}{\xi}}\label{eq_thetaasymptotic},
\end{align}
where $\mathcal{K}_n$ is the modified Bessel function of the second kind.
The healing length,  
\begin{align}
    \xi = \frac{1}{\kappa}\sqrt{\frac{\Bs}{B+2K}}\label{eq_xidef},
\end{align}
sets the scale for decay of the in-plane magnetization far from the skyrmion.
The asymptotic solution given in Eq.~\eqref{eq_thetaasymptotic} can be further
simplified to
$\Theta\qty(r)\propto\sqrt{\flatfrac{\xi}{r}}\exp(-\flatfrac{r}{\xi})$
in the limit $r\gg \xi$.  The skyrmion radius, $R$, may be found numerically by inverting 
\begin{align}
    \Theta\qty(R) = \frac{\pi}{2}\label{eq_Rdef}
\end{align}
for $\Theta\qty(r)$ satisfying Eq.~\eqref{eq_1skODE}.
At the skyrmion radius, $r=R$, the magnetization is entirely in-plane.
The magnetic texture of an isolated skyrmion is plotted in
Fig.~\ref{fig_1sk_cmaps}\!(a). 

Although Eq.~\eqref{eq_FMfdv} has a skyrmion solution for any finite $B$ or $K$, this solution does not describe a minimum in the free energy for sufficiently small $B$ and $K$. Below a critical line in the $B,K$ plane, skyrmions are unstable to a transition towards a spin-spiral phase.\cite{2017_Han} We numerically determine this phase boundary, plotted in Fig.~\ref{fig_magnetic_parameters}(a) (see also Table \ref{table_nparams}) through
the procedure described in Appendices~\ref{ap_2sknumerics} and ~\ref{ap_stability}. Above this phase boundary, localized skyrmions exist as stable minima of the magnetic free energy. A more extensive review of the theory of magnetic skyrmions is given in Refs.~[\onlinecite{2013_Nagaosa, 2017_Han}].

\begin{figure}
    \includegraphics{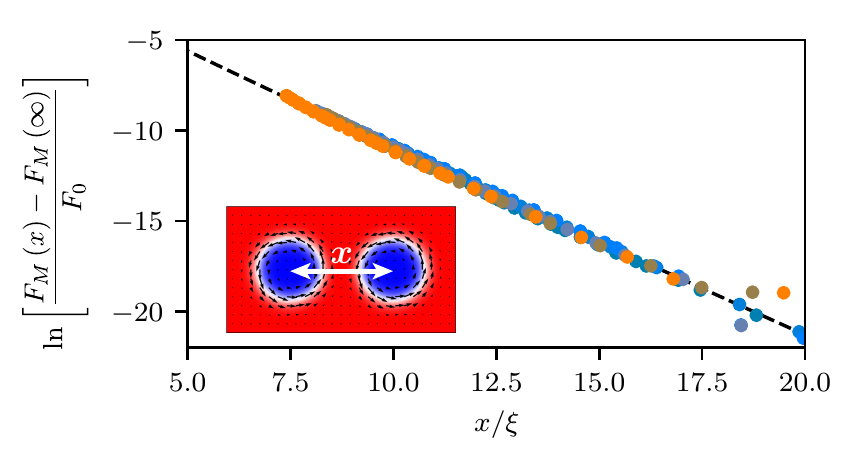}
    \caption{The dependence of the magnetic free energy, $\FM$, on skyrmion-skyrmion
    separation, $x$, for a pair of skyrmions. The magnetic free energy was determined
    numerically by pinning the position of the skyrmions and allowing the
    remaining magnetic degrees of freedom to relax (see
    Appendix \ref{ap_2sknumerics} for details). We compare $\FM(x)$ for several
    values of the applied field, $B$, and anisotropy, $K$ (colored dots). The single-skyrmion
    length scales for each $\qty(B,K)$ as well as the color-scheme are presented
    in Table~\ref{table_nparams}.  The magnetic free energy relative to its large-separation limit, $\FM\qty(x)-\FM(\infty)$, is fit with the
    function ${\BesselK{1}{\flatfrac{x}{\xi}}}{}$ (dashed line) and the proportionality factor $F_0$ is determined independently for each data set. The healing length, $\xi$, is set by Eq.~\eqref{eq_xidef}. 
    Inset: the magnetization profile for $B/B_s=0.1$, $K/B_s=0.9$ with $x/\xi=9.12$.}\label{fig_FMofX}
\end{figure}

We now consider a system of two skyrmions separated by a distance 
$x$. Our goal is to establish that the intrinsic inter-skyrmion interaction
is repulsive and short-range, with length scale $\xi$. This form of the
intrinsic  interaction is expected based on the following reasoning. Far from
the skyrmion core, the magnetization is mostly aligned with the applied field
save for a small in-plane component. Two skyrmions repel each other because
they favor opposing in-plane magnetization in the region between
them.\cite{2013_Lin} The magnitude of this in-plane component is determined by
the length scale $\xi$. We therefore expect a short-range repulsive interaction
between skyrmions with length scale $\xi$.

To verify this picture, we treat this problem numerically on a lattice in a
similar manner to Ref.~[\onlinecite{2013_Lin}]. We prepare an initial magnetic configuration with two skyrmions separated by a specified distance.  We then
freeze a magnetic moment within each skyrmion core and minimize the magnetic
free energy with respect to variations in the remaining moments. Given this relaxed
magnetization, we calculate the magnetic free energy and establish the final
inter-skyrmion separation, $x$ for the relaxed configuration. By repeating this calculation with different initial inter-skyrmion separations, we determine the magnetic free energy as a
function of $x$. Further details are given in Appendix \ref{ap_2sknumerics}.
The results of this calculation are given in Fig.~\ref{fig_FMofX}.  We find that
the intrinsic magnetic inter-skyrmion interaction is well fit by
a modified Bessel function (of first order and second kind) with the separation, $x$,
scaled by $\xi$, consistent with Eq.~\eqref{eq_thetaasymptotic}. This agreement holds over several orders of magnitude of
interaction strength. For sufficiently large inter-skyrmion separations,
errors in the numerical calculation of $\FM(x)$ become comparable to the
interaction energy [$\FM(x)-\FM(\infty)$], and we can no longer make numerical predictions. These
results are entirely consistent with the analogous calculation presented in
Ref.~[\onlinecite{2013_Lin}], where it was shown that the magnetic healing length
determines the range of the repulsive skyrmion-skyrmion interaction for $K=0$.

In the analysis given above, we have neglected the long-range
magnetic dipolar interaction. This can be justified
for skyrmions with a sufficiently small radius $R$ (reducing the total moment
associated with each skyrmion) and for a moderate separation $x$. 
In Sec.~\ref{sec_considerations}, we provide a more detailed justification of this approximation in the context of a candidate material, $\textrm{Cr}_{x}{(\textrm{Bi}_{1-y}\textrm{Sb}_{y})}_{2-x}\textrm{Te}_{3} /
{(\textrm{Bi}_{1-y}\textrm{Sb}_{y})}_{2}\textrm{Te}_{3}$.

\definecolor{c1}{rgb}{0.0,0.5,0.6923076923076922}
\definecolor{c2}{rgb}{0.0,0.5,0.846153846153846}
\definecolor{c3}{rgb}{0.0,0.5,1.0}
\definecolor{c4}{rgb}{0.4000000000000001,0.5,0.6923076923076922}
\definecolor{c5}{rgb}{0.6000000000000001,0.5,0.29230769230769227}
\definecolor{c6}{rgb}{1.0,0.5,0.0}

\begin{table}
    \begin{ruledtabular}
        \begin{tabular}{p{.2\columnwidth}p{.2\columnwidth}p{.2\columnwidth}p{.2\columnwidth}p{.2\columnwidth}}
            Color&$\kappa\xi$&$\kappa R$&$B/\Bs$&$K/\Bs$\\
\textcolor{c1}{\Large $\bullet$} & 0.73 & 2.85 & 0.1 & 0.9\\
\textcolor{c2}{\Large $\bullet$} & 0.69 & 2.35 & 0.1 & 1.0\\
\textcolor{c3}{\Large $\bullet$} & 0.66 & 1.94 & 0.1 & 1.1\\
\textcolor{c4}{\Large $\bullet$} & 0.71 & 1.84 & 0.2 & 0.9\\
\textcolor{c5}{\Large $\bullet$} & 0.81 & 2.14 & 0.25 & 0.64\\
\textcolor{c6}{\Large $\bullet$} & 0.89 & 2.14 & 0.35 & 0.45
        \end{tabular}
    \end{ruledtabular}
    \caption{Magnetic-free-energy parameters and single-skyrmion length scales
    for the numerical  results used throughout this work. The single-skyrmion length scales, $\xi$ and $R$, are
    determined from $B$ and $K$ through Eqs.~\eqref{eq_xidef}
    and~\eqref{eq_Rdef}. We choose $B$ and $K$ within the stable region of the
    skyrmion phase diagram [see Fig.~\ref{fig_magnetic_parameters}(a)] to tune
    the ratio $\xi/R$. 
    We set $\lambda_0=R$ when calculating the electronic states to ensure that the skyrmions host discrete sets of orbitals.}\label{table_nparams}
\end{table}

\section{Attractive interaction}
\label{sec_esystem}
This section addresses the attractive interaction effected by the electronic system. First, we consider 
the electronic structure of the MTI surface resulting from the hybridization of
single-skyrmion orbitals. Next, we integrate out the electronic subsystem,
accounting for an electronic reservoir at fixed chemical potential, to determine
the grand potential as a function of skyrmion-skyrmion separation. This analysis
leads us to conclude that the MTI surface states can give rise to an attractive
skyrmion-skyrmion interaction.

In the absence of a magnetic subsystem, a three-dimensional topological
insulator surface is characterized by an odd number of spin-orbit-coupled Dirac
cones. We consider a single Dirac cone centered at $\k = 0$, which is coupled to
the magnetic system through a Zeeman-like term proportional to the local
magnetization:
\begin{multline}
    H_\mathrm{MTI}= \hbar v\sum_{\bf k}c_{\bf k}^\dagger  {\qty(\k\times\bsigma)}_{z} c_{\bf k}  \\
    -\Delta\int d^2 {\bf r} \psi_{\bf r}^\dagger \m\qty(\r)\cdot\bsigma \psi_{\bf r},
    \label{eq_MTIHamiltonian}
\end{multline}
where $c_{\bf k}$ is a spinor of electronic annihilation operators in momentum and $\psi_{\bf r}  = {1\over \sqrt A} \sum_{\bf k} \exp(i{\bf k\cdot r }) c_{\bf k}$ is its real space counterpart defined with the system's area $A$. Furthermore, $v$ is the Fermi velocity, and $\Delta$ is proportional to the exchange
interaction between the surface electrons and the magnetic system. These
parameters imply a natural length scale:
\begin{align}
    \lO = \frac{\hbar v}{\Delta},\label{eq_l0defn}
\end{align}
and a natural energy scale, $\Delta$.

The MTI surface Hamiltonian, Eq.~\eqref{eq_MTIHamiltonian}, neglects the magnetic vector potential. This is justified if the $U(1)$ phase acquired
by electrons in the skyrmion-bound orbitals is small, i.e., the flux through the skyrmion area is smaller than the flux quantum.  This condition can be written as $R^2\tilde{B} \ll
\Phi_0$ where $\tilde{B}\propto B$ is the dimensionful applied
magnetic field. For a skyrmion of radius
$R=50$~nm, this condition is satisfied for $\tilde{B} \ll 1$ T.
Equation~\eqref{eq_MTIHamiltonian} also neglects the
direct Zeeman coupling of the surface electrons to the applied magnetic field.
This is justified if the Zeeman-like exchange coupling to the magnetization is
large compared to the Zeeman energy. In Sec.~\ref{sec_considerations}, we show
that this limit is experimentally relevant.

For a uniform magnetization and finite $\Delta$, $\mz$ gives the surface Dirac electrons a finite mass, while the in-plane components simply shift the Dirac point from $\k=0$. The resulting massive Dirac cone is characterized by a finite Berry curvature, with sign determined by the sign of $m_z$. Thus, gapless states will be found where $m_z$ changes sign.\cite{1976_Jackiw,1984_Jackiw} These states are given below as the eigenstates of Eq.~\eqref{eq_MTIHamiltonian} with a magnetization
given by the solution of Eq.~\eqref{eq_1skODE}. 
Since the skyrmion magnetization has rotational symmetry, the energy eigenstates are a product of angular and radial functions:
   \begin{align}
       \braket{\r}{\psi_{j}} = \mqty(e^{i\qty(j-\half)\phi} & 0 \\
       0& e^{i\qty(j+\half)\phi})\boldsymbol{\chi}_j\qty(r)\label{eq_totalwavefn},
   \end{align}
where $j$ is the half-integer total-angular-momentum quantum number, and $\boldsymbol{\chi}_j\qty(r)= \qty(\chi_j^\uparrow\qty(r),\chi_j^\downarrow\qty(r))^{T}$ is the radial wave function.
If the in-plane component of the magnetization is divergenceless, as in the case of Bloch skyrmions, we may work in a gauge where it does not enter the radial wave equation:\cite{2015_Hurst}
\begin{align}
    \mqty(-\mz\qty(r)& -\lO\dv{r}-\lO\frac{j+\half}{r} \\
    \lO\dv{r}-\lO\frac{j-\half}{r} & \mz\qty(r))
    \boldsymbol{\chi}_j\qty(r)
    =\frac{E_{j}}{\Delta}
    \boldsymbol{\chi}_j\qty(r).
    \label{eq_radial_wave_eqn}
\end{align}
For ${r-R}{}\gg \xi$, the magnetization approaches $\zh$ and the asymptotic
behavior of the bound-state wavefunction is 
\begin{align}
    \boldsymbol{\chi}_j\qty(r)
    \sim
    \mqty(\sqrt{1-\frac{E_j}{\Delta}}\BesselK{j-\half}{\frac{r}{\lambda_j}}\\
    \sqrt{1+\frac{E_j}{\Delta}}\BesselK{j+\half}{\frac{r }{\lambda_j}}),\label{eq_1skasymptotic}
\end{align}
where 
\begin{align}
    \lambda_j = \frac{\lO}{\sqrt{1-{\qty(\frac{E_j}{\Delta})}^2}}.\label{eq_lambdadefn}
\end{align}

For $R \gtrsim \lO$ there is a discrete set of states within the Dirac mass gap
with $\bra{\r}\ket{\psi_j}$ localized to the skyrmion. Away from the skyrmion,
these wavefunctions decay with length scale $\lambda_j$.
Hurst {\it et al.} (Ref.~\onlinecite{2015_Hurst}) provide an explicit solution for the wavefunctions in the limit $\xi=0$ as well as a transcendental equation for the
electronic spectrum as a function of $R$ (similar solutions are also given for related models in Refs.~\onlinecite{2013_Ferreira,2015_Uchoa}). We determine the
bound states exactly on a lattice for the full single-skyrmion magnetization,
$\m^\mathrm{sk}\qty(\r)$. The magnetization profile, $\m^\mathrm{sk}\qty(\r)$, is determined from the solution to
Eq.~\eqref{eq_1skODE} using the approach discussed in
Appendix \ref{ap_electronic}. In Fig.~\ref{fig_1sk_spectrum}, we plot the spectrum
from Ref.~[\onlinecite{2015_Hurst}] as a function of $R$ (black lines) together with the
numerically determined spectrum for a skyrmion with $\xi=0.256\,R$ (blue dots).

\begin{figure}
    \includegraphics{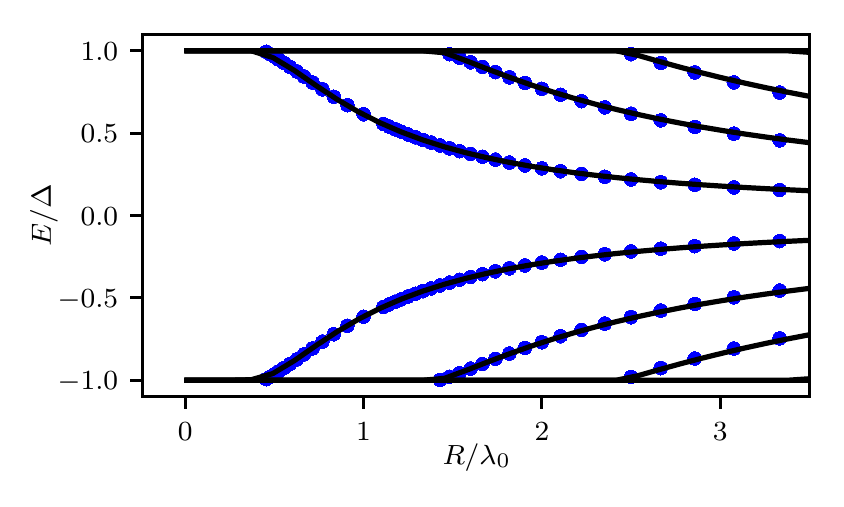}
    \caption{Dependence of the in-gap spectrum on skyrmion radius. The
    spectrum given in Ref.~[\onlinecite{2015_Hurst}] for $\xi=0$ (black lines) is plotted with the 
    numerically determined spectrum for $B/\Bs=0.1$ and $K/\Bs=0.9$  (blue dots). For
    $R/\lO\sim1$, the skyrmion hosts a discrete set of bound
    orbitals, while for $R\gg\lO$ the skyrmion hosts a linearly dispersing
    continuum of chiral edge states.}\label{fig_1sk_spectrum}
\end{figure}

In a system hosting two skyrmions, the single-skyrmion orbitals will overlap and
hybridize to form molecular orbitals. We focus on the lowest-energy electronic state with $j=1/2$ in each skyrmion and construct a two-level Hamiltonian to describe the orbital hybridization:
\begin{align}
    H^\textrm{Mol} = \sum_{i=1,2}E_0 \cid\ci + t\qty( c_1^\dagger c_2 + c_2^\dagger
    c_1)\label{eq_hopping},
\end{align}
where $E_0$ is the energy of the lowest-energy single-skyrmion orbital and $c_i$ annihilates an electron in the lowest-energy single-skyrmion orbital of skyrmion $i$. The
tunneling amplitude is given by 
\begin{align}
    t = \mel{\psi^1}{M_1}{\psi^2} = \mel{\psi^2}{M_1}{\psi^1}\label{eq_tdefn},
\end{align}
where $\ket{\psi^i}$ are the lowest-energy ($j=\flatfrac{1}{2}$ for $R=\lO$ and $v>0$) single-skyrmion orbitals centered at $\r_i$, defined through
Eq.~\eqref{eq_totalwavefn}, and 
\begin{align}
    M_{i} = -\Delta \qty[\m^\mathrm{sk}\qty(\r-\r_{i})-\zh]\cdot\bsigma\label{eq_1skM}
\end{align}
is the `atomic potential' of skyrmion $i$ relative to the ferromagnetic
background. The tunneling amplitude, $t\qty(x)$, 
decays exponentially for ${x-2R}{}\gg\l$ where $\l \equiv \l_j$ for $j$ corresponding to the lowest-energy single-skyrmion orbital. This parametric dependence comes from the
asymptotic behavior of the single-skyrmion orbitals defined in
Eq.~\eqref{eq_1skasymptotic}.

\begin{figure}
    \includegraphics{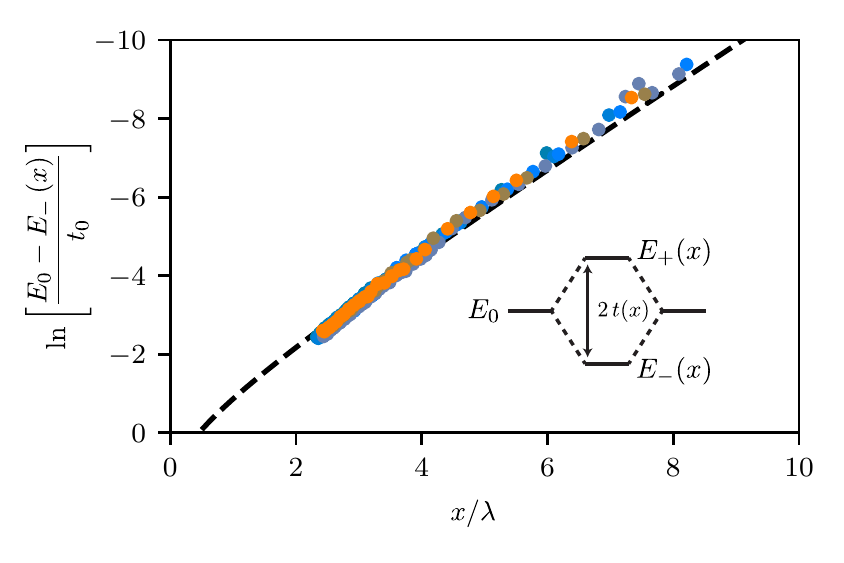}
    \caption{Molecular binding energy relative to the energy of the
    single-skyrmion $j=\flathalf$ state as a function of skyrmion-skyrmion
    separation. The single-skyrmion orbitals decay into the magnetic bulk with
    length scale $\lambda$ [Eq.~\eqref{eq_lambdadefn} with $j=\flathalf$],
    resulting in a similar decay for $t(x)$. 
    The phenomenological dependence
    considered in Sec.~\ref{sec_phasediagram}, ${E_{0}-E_-}{{}}~=~{t_0}
    \BesselK{0}{\frac{x}{\lambda}}$, is plotted with the dashed line. Here, $t_0$ is determined for each data set independently
    through a fit to the numerical results (colored circles, corresponding to the parameters given in Table~\ref{table_nparams}, which are determined using the approach described in Appendix \ref{ap_electronic}). The single-skyrmion orbital decay length, $\l$, is calculated using Eq.~\eqref{eq_lambdadefn} with the numerically determined energy of the single-skyrmion orbitals.}\label{fig_hybridization}
\end{figure}
Equation~\eqref{eq_hopping} is diagonalized by molecular orbitals
with energies
\begin{align}
    E_{\pm} = E_0\pm\abs{t\qty(x)}.\label{eq_molecular_energies}
\end{align}
We neglect the doubly-occupied state for a skyrmion pair at a sufficiently small
separation $x$ and for a weakly-screened long-range Coulomb interaction.
In Sec.~\ref{sec_considerations}, we justify this approximation in the context of recently synthesized Cr doped ${\qty(\mathrm{Bi}_{2-y}\mathrm{Sb}_{y})}_{2}\mathrm{Te}_3$ heterostructures.

Figure~\ref{fig_hybridization} presents the binding energy as a function of $x$
for skyrmions with radius $R = \lambda_0$.  The binding energy is the difference
between the lowest-lying molecular orbital, $E_-\qty(x)$, and the
single-skyrmion orbital energy, $E_0 = E_-\qty(\infty)$.  The binding energy indeed decays with
length scale $\lambda$, and is well fit by
$t_0\BesselK{0}{\frac{x}{\lambda}}$, as expected from Eq.~\eqref{eq_tdefn},
and  the asymptotic single-skyrmion wavefunctions, given in
Eq.~\eqref{eq_1skasymptotic}.

\begin{figure}
    \includegraphics{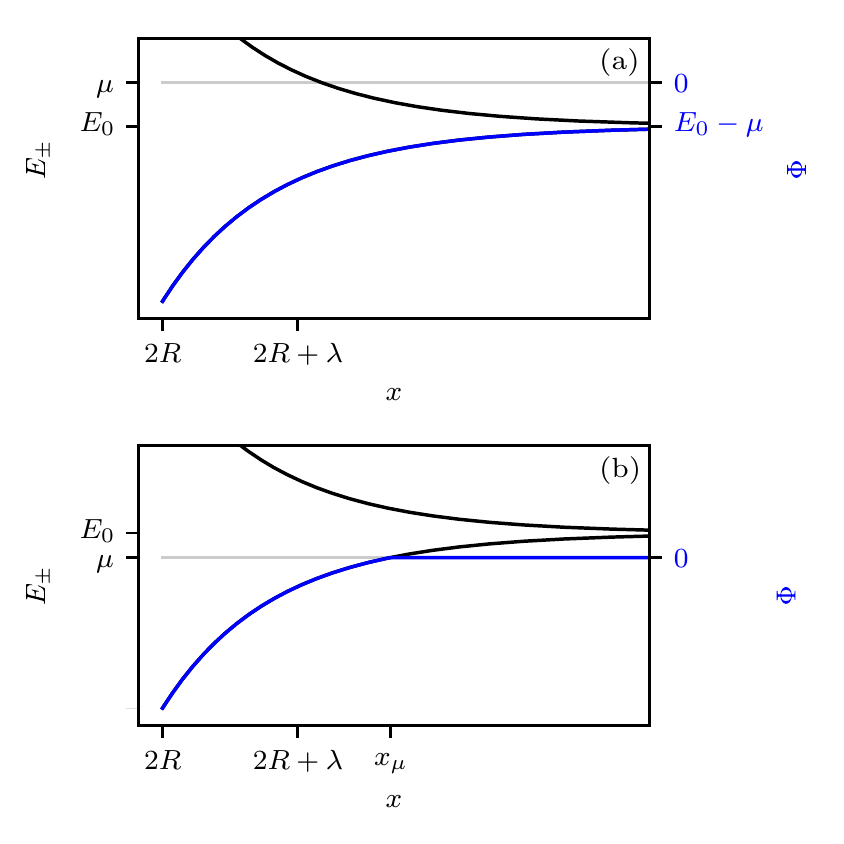}
    \caption{Two-skyrmion electronic spectrum ($E_+$ and $E_-$, black) and zero-temperature
    grand potential ($\Phi$, blue) at (a) $\mu > E_0$ and (b) $\mu < E_0$  for
    the tunneling amplitude given in Eq.~\eqref{eq_phenomt}.
    The chemical potential, $\mu$, is plotted in gray. 
    For $\mu>E_-$, the lowest-energy molecular orbital is occupied and the grand
    potential is given by $\Phi=E_--\mu$.
    When $\mu<E_-$, the lowest-energy molecular orbital is unoccupied and the
    grand potential is simply given by the energy of the unoccupied state
    ($\Phi=0$).
    The tunneling amplitude, $t=\flatfrac{\qty(E_+- E_-)}{2}$, 
    decays exponentially with separation, over the single-skyrmion orbital decay
    length, $\lambda$.}\label{fig_grandpotential}
\end{figure}

The effective interaction between a skyrmion pair is determined by integrating out the electronic degrees
of freedom.  The electronic state is determined through the grand potential for a fixed separation $x$:  
\begin{align}
    \Phi\qty(x) =& -T\ln\qty(\Z),\label{eq_GPdefn}\\
    \Z =& 1 + e^{-\frac{E_- - \mu}{T}}+ e^{-\frac{E_+ - \mu}{T}},\label{eq_Zdefn}
\end{align}
Where we set Boltzman's constant to 1 and assume that the electronic subsystem is in contact with a reservoir at temperature $T$ and chemical potential $\mu$. The molecule is restricted to support $N=0,1$ electrons, consistent with the discussion
following Eq.~\eqref{eq_molecular_energies}. 

The $T=0$ behavior of $\Phi$ is sketched in Fig.~\ref{fig_grandpotential}. For
$\mu > E_0$, the lowest-energy molecular orbital is occupied at all $x$ and the
grand potential is $\Phi = E_-\qty(x)-\mu$. For $\mu < E_0$, the lowest-energy
molecular orbital crosses the chemical potential at a separation $\xmu$, above
which it becomes unoccupied. $x_\mu$ is defined by
\begin{align}
    E_-\qty(\xmu) \equiv \mu\label{eq_xmudefn}.
\end{align}
For a monotonically increasing energy $E_-\qty(x)$ at $T=0$, the grand potential is therefore:
\begin{align}
    \Phi  = 
    \begin{cases} 
        E_-\qty(x)-\mu & x < \xmu\\
        0 & x > \xmu 
    \end{cases}. \label{eq_Phipiecewise}
\end{align}

Since $-\Phi^\prime(x)< 0$, at $T=0$, the electronic system effects an attractive force
for $x<\xmu$ or when $\mu > E_0$. If this attractive
interaction overcomes the short-range repulsive magnetic interaction
discussed in Sec.~\ref{sec_msystem}, a stable bound skyrmion molecule will form.

\section{Pair-binding phase diagram}
\label{sec_phasediagram}
The total free energy is given by the sum of the contributions from the
electronic system and the magnetic system,
\begin{equation}
    F(x) = F_M(x)+\Phi(x).\label{eq_FofX}
\end{equation}
Based on the possible functional forms of $F(x)$ we identify four regimes whose nature and boundaries will be discussed below.
The phase diagram is depicted in Fig.~\ref{fig_phasediagram}(a) and a representative free-energy plot for each phase is shown in Figs.~\ref{fig_phasediagram}(b)-(e).
We denote the regimes of our model as:\\
(i) the `stable' regime, where the free energy has a unique global minimum at finite skyrmion-skyrmion separation. 
We denote this separation $x_B$ and refer to it as the bond length.
The stable regime is colored green in Fig.~\ref{fig_phasediagram}.\\
(ii) the `stable-hysteretic' regime, where, in addition to the global minimum at $x_B$, there is a local minimum as $x\to\infty$.
This regime is colored in beige in the Fig.~\ref{fig_phasediagram}.\\
(iii) the `metastable-hysteretic' regime, where there is a \emph{local} minimum at $x_B$ and a \emph{global} minimum at infinite separation.
This phase is blue in Fig.~\ref{fig_phasediagram}.\\ 
(iv) the `unstable' regime, where the free energy is a monotonically decreasing function of separation, and there is no bound state.
This phase is colored white in Fig.~\ref{fig_phasediagram}. 

\subsection{Zero temperature}
At zero temperature, the electronic free energy is defined piecewise through $\Phi=\min\left[E_-(x)-\mu, 0\right]$, as shown in Eq.~\eqref{eq_Phipiecewise} and Fig.~\ref{fig_grandpotential}(b).
For any pair of $\mu$, $\xi$, the phase at zero temperature is determined from the balance of the repulsive (magnetic) and attractive (electronic) forces at a finite separation.
We recall that the magnetic contribution to the free energy, $F_M$, is always repulsive and decays quickly over a length scale $\xi$, while the electronic contribution is attractive whenever the lower orbital state is populated and is governed by the length scale $\lambda$.
Moreover, we note that, for a bound state to form, the repulsion should be shorter-range than the attraction, $\xi<\lambda$. 
We therefore scan $\xi<\lambda$ and find the phase boundaries for each $\xi$ as a function of the chemical potential, $\mu$.

\begin{figure}[!] 
 \includegraphics{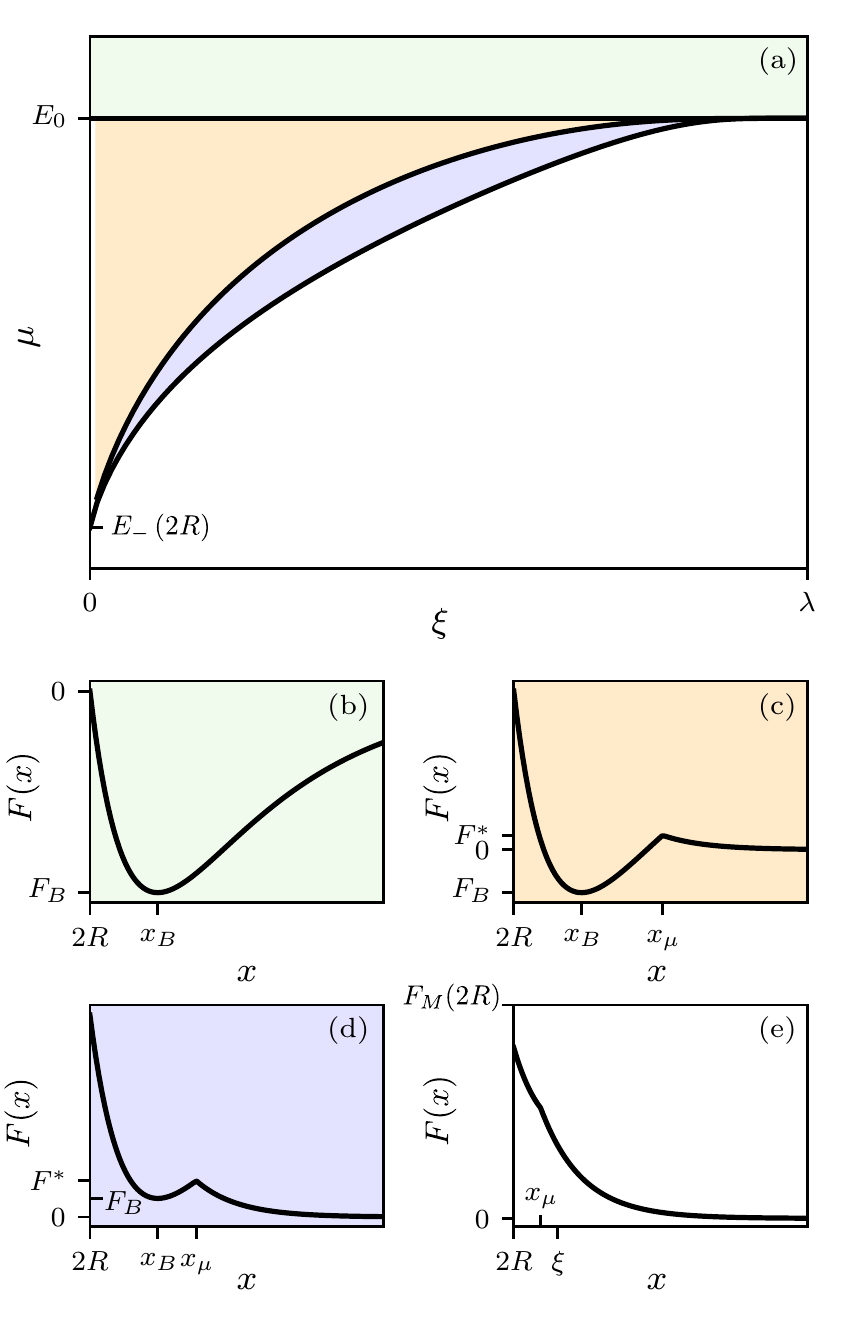} 
 \caption{(a) Zero-temperature phase diagram describing regions of stable (green), stable-hysteretic (beige), metastable-hysteretic (blue), and unstable (white) skyrmion pair binding as a
 function of the chemical potential, $\mu$, and magnetic healing length, $\xi$,
 for $\xi<\lambda$.
    (b) For large chemical potential ($\mu >E_{0}$) the
 lowest-energy molecular orbital is occupied and the total skyrmion-skyrmion
    interaction, $F = \Phi+F_M$, has a unique minimum, $F\qty(x_B)$, at a finite
 skyrmion-skyrmion separation, $x_B$.
    Thus, the skyrmions form a bound state.
 (c,d) For intermediate chemical potential, the lowest-energy molecular
 orbital is occupied for $x<x_\mu$, but is unoccupied for $x>x_\mu$.
    The bound
 configuration remains locally stable, but for $x>x_\mu$ the skyrmion-skyrmion
 interaction is repulsive.
    Thus, both the bound and unbound configurations will be long-lived.
    Which of these configurations is the ground state is determined by comparing $F\qty(x_B)$ and $F\qty(x\to\infty)$.
    (e) For sufficiently low chemical potential, we have $x_\mu < x_b$, and the
 skyrmion-skyrmion interaction is strictly repulsive.
    The unbound
 configuration is then favored.  Please note that we set the unknown $F_0$ such that $F(2R)=t_0(2R)$ for each $\xi$.~\label{fig_phasediagram}} 
\end{figure}

\emph{(i) The stable phase}---if $\mu$ is above the lowest molecular orbital energy, $E_-$, for all separations $x$, i.e., $\mu>E_0$, then the orbital is always occupied and $\Phi(x) = E_-(x)-\mu$ for all $x$.
We therefore have both repulsion and attraction everywhere and a balance of forces occurs at $x_B$ satisfying
\begin{equation}\label{eq_xb}
    F_M'(x_B) + E_-'(x_B) = 0.
\end{equation}  
This stable phase is characterized by a free energy with a unique minimum at $x_B$. 
In Fig.~\ref{fig_phasediagram}, the regime is bounded from below by
\begin{equation}
\mu>E_0.
\end{equation}

\emph{(ii) The stable-hysteretic regime}---in addition to the stable regime, there are two other regimes in which a bound state occurs.
As discussed above, when the chemical potential crosses $E_-(x)$ at some separation $x_\mu$, there is an attractive force for $x<x_\mu$ but no attraction for $x>x_\mu$.
If the forces balance at a separation where the lower-energy orbital is populated, i.e., $x_B<x_\mu$, there will be a skyrmion bound state.
The stable-hysteretic regime and the stable regime are distinguished by the behavior of the free energy at large separations.
Because there is no attractive interaction for $x>\xmu$ in the stable-hysteretic regime, there is a free-energy minimum as $x\to \infty$.
Between the to minima there is a cusp-like energy barrier at $x_\mu$ as depicted in Fig.~\ref{fig_phasediagram}(c).
In the stable-hysteretic phase the minimum at $x_B$ arises at a lower free energy than the minimum at $x\to\infty$, 
\begin{equation}
F(x_B)<F(x\to\infty),
\end{equation}  
so the bound state is stable.
However, two skyrmions prepared at large separation experience a repulsive interaction and so remain unbound.
This is the hysteretic behavior alluded to in the name of the phase.

\emph{(iii) The metastable-hysteretic regime}---the metastable-hysteretic regime is distinguished from the stable-hysteretic regime by the relative energies of the bound and unbound configurations.
In this regime the unbound configuration globally minimizes the free energy,
\begin{equation}
F(x_B)>F(x\to\infty),
\end{equation}  
but the molecular orbital is occupied at the bond length, $x_B<x_\mu$.
The free-energy minimum at $x_B$ is local and the bound state is therefore metastable.
The lower bound of this phase is found when the molecular orbital is depopulated at $x_B$, $x_B=x_\mu$, and the attraction can no longer overcome the repulsion at any separation. 

\emph{(iv) The unstable regime}---in this regime, the repulsive interaction dominates the free energy for all separations. 
In this case the only free-energy minimum occurs as $x\to \infty$.
This occurs when the molecular orbital is unoccupied at $x_B$, as defined in Eq.~\eqref{eq_xb}, so $x_B$ is not a minimum of the free energy. 
The unstable phase is defined by the inequality:
\begin{equation} 
x_\mu < x_B,
\end{equation}
and the phase boundary is found by comparing the two lengths.

In Fig.~\ref{fig_phasediagram}(a), we estimate the phase boundaries as described above using the asymptotic behaviour of $F_M(x)$ and $\Phi(x)$.
In Fig.~\ref{fig_FMofX}, we show that for $x\gg\xi,2R$, the magnetic free energy, $F_M(x)$ has the asymptotic form
\begin{equation}
F_M(x) \sim F_0 \BesselK{1}{\frac{x}{\xi}},
    \label{eq_phenomFMofx}
\end{equation}
where $F_0$ characterizes the strength of the repulsive interaction.
Similarly, in Fig.~\ref{fig_hybridization}, we show that the tunneling
amplitude, $t(x)$ at $x\gg\l$, has the asymptotic form
\begin{equation}
t(x) \sim t_0 \BesselK{0}{\frac{x}{\l}}\label{eq_phenomt},
\end{equation}
where $t_0$ characterizes the strength of the tunnel splitting.

We use the above asymptotic forms to find the bond length $x_B$, the scale $x_\mu$,  and the free energy at the minimum $F(x_B)$ and use these quantities to draw the phase diagram in Fig.~\ref{fig_phasediagram}.
Moreover, it is possible to approximate the functional form of the free energy further by taking the asymptotic limit of the above Bessel functions:
\begin{align}
    F_M(x) &\sim\sqrt{\pi\over 2} F_0 \left({x\over \xi}\right)^{-{1\over 2}} e^{-x/\xi} && x\gg\xi, 
\label{eq_approxFM}\\
    E_-(x) &\sim E_0-\sqrt{\pi\over 2} t_0 \left({x\over \lambda}\right)^{-{1\over 2}} e^{-x/\lambda} && x\gg\lambda.
\label{eq_approxE-}
\end{align}
Balancing the derivatives of the two expressions above and neglecting subleading terms in $\lambda/x$ and $\xi/x$, we obtain:
\begin{eqnarray}
    x_B \sim \frac{\lambda\xi}{\lambda-\xi} \ln\left(\sqrt{\frac{\lambda}{\xi}}\frac{F_0}{t_0}\right).\label{eq_xBasym}
\end{eqnarray}
This limit is consistent with our assumption of $x \gg \lambda >\xi$ when 
\begin{eqnarray}
{F_0 \over t_0} \ll \sqrt{\xi\over \lambda}.
\end{eqnarray}
Equation~\eqref{eq_xBasym} shows that the bond length becomes shorter as the ratio $\xi/\lambda$ is decreased.
The minimum of the free energy can be found by substituting Eq.~\eqref{eq_xBasym} into the approximated form of $F(x)$ using Eqs.~\eqref{eq_phenomFMofx}, \eqref{eq_phenomt} or Eqs.~\eqref{eq_approxFM}, \eqref{eq_approxE-}.
For $x\gg \xi,\lambda$, $E_-(x)$ and $F_M(x)$ are approximately exponential giving
 \begin{eqnarray}
     F_M'(x) \sim -{1\over \xi} F_M(x) && E_-'(x)\sim -{1\over\lambda}t(x).
 \end{eqnarray}
This helps simplify the expression for $F(x_B)$ as the forces [$-F_M'(x)$, $-E_-'(x)$] are equal in magnitude and opposite in sign at the minimum $x_B$.
We may therefore write
 \begin{eqnarray}
     F(x_B)& = E_-(x_B)+F_M(x_B)-\mu,\\
     &\sim E_-(x_B)\left(1-\frac{\xi}{\lambda}\right) -\mu,
 \end{eqnarray}
which demonstrates that, as expected, the bound-configuration minimum is deeper for smaller ratios of $\xi/\lambda$.

 One should note that these estimates are based on the asymptotic behavior of $t(x)$ and $F_M(x)$ and hence are accurate only in the limits mentioned above.
 However, we stress that a bound state may exist well beyond these limits.
 For example, in the limit where $\xi \ll \lambda$ we may view the magnetic repulsion as a hard-shell repulsion.
 Therefore, the bound state would form at $x=2R$ where the repulsion drops to zero while the attractive force is given by $-E_-'(2R)\ne 0$.

\subsection{Finite temperature}
\begin{figure}
	\includegraphics[width=\columnwidth]{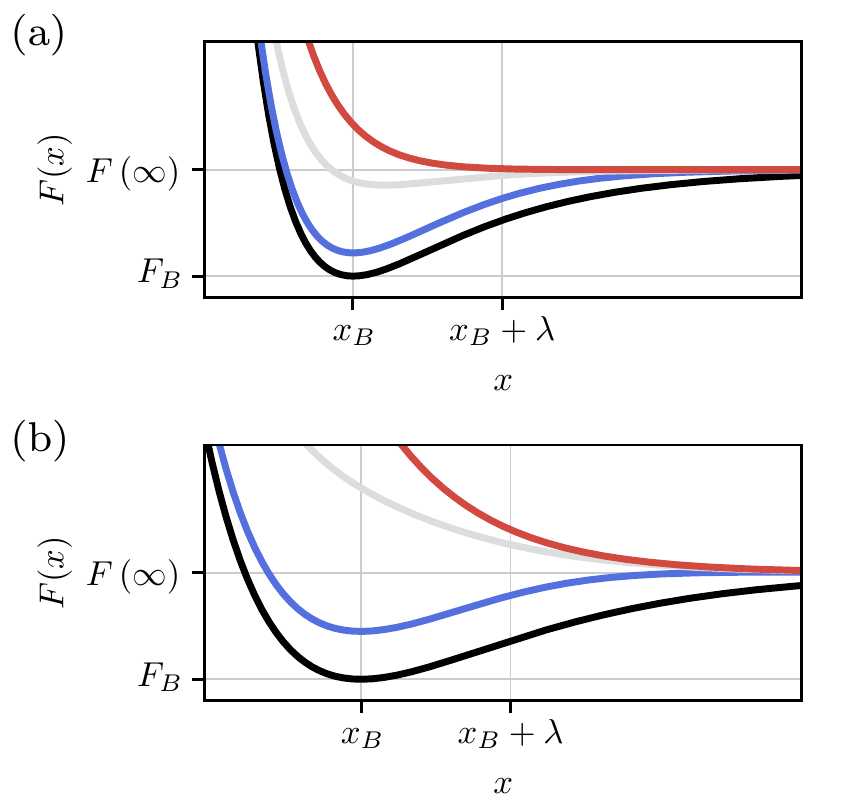}
    \caption{Total free energy as a function of separation at $T=0$ (black), $T=\flatfrac{T^*}{5}$ (blue), $T=T^*$ (grey), and $T=5T^*$ (red).
    (a) For $\xi=\lambda/3$, there is a well-defined free-energy minimum for all $T$, but this minimum is suppressed for $T> T^*$. 
    (b) For $\xi = 2\lambda/3$, there is a free-energy minimum only for $T\ll T^*$.\label{fig_finite_T}}
\end{figure}

At finite temperature the free energy changes due to thermal occupation of the molecular orbital states as well as thermal fluctuations in the skyrmion separation $x$ and possible changes in the magnetic free energy.
In this section we include the temperature only through its effect on the population of the molecular orbitals.
Beginning in the stable regime at zero temperature, we determine the conditions for which there may still be a bound state when the temperature is raised.
This may be addressed qualitatively by plotting $F(x)$ as defined in Eq.~\eqref{eq_FofX}, with the finite-temperature $\Phi$ given by Eq.~\eqref{eq_GPdefn}, see, e.g.,  Fig.~\ref{fig_finite_T}.

We assess the effect of temperature on the bound state in the following way.
We assume a large chemical potential such that one of the molecular orbitals is always occupied, $\mu-E_\pm\gg T$ (but we still neglect double occupancy).
At finite temperature, the average occupations of the states with energies $E_-$ and $E_+$ shift away from the zero-temperature limits of 1 and 0.
The electronic free energy, $\Phi(x)=\bar{E}-TS$, has contributions from both the entropy $S$ and the average energy $\bar{E}$.  When $T\gg E_+-E_-$, the populations of the two states are comparable, giving $S\to \ln 2$, and the average energy also becomes independent of $x$: $\bar{E}\to \left(E_+(x)+E_-(x)\right)/2=E_0$.  Since $\Phi$ becomes $x$-independent, the attractive force is suppressed as the temperature is raised.
We therefore define the temperature scale $T^*$, controlled by the typical molecular energy scale at the free-energy minimum:
\begin{equation}
T^* = t(x_B).
\end{equation}
For $T\ll T^*$, the population of the $E_-$ state is significantly larger than that of $E_+$, the zero-temperature analysis applies, and the bound state persists.
Above this temperature, the electronic contribution to the free energy is suppressed.
For $T>T^*$ the suppression of the attractive force is linear in $t(x)/T$ and therefore one may find a free-energy minimum even above $T^*$.
Since one of the molecular states is always occupied under the conditions described above, we can write the partition function
\begin{eqnarray}
\Z = e^{E_0-\mu}\left(e^{-t(x)/T}+e^{t(x)/T}\right),
\end{eqnarray}
and consequently find the attractive force,
\begin{eqnarray}
f(x) = {T\over \Z}{d \Z \over dx} = \tanh\left({t(x)\over T}\right)t'(x). 
\end{eqnarray}
At high temperature, the attractive force $f(x)$ is suppressed, and decays exponentially at half the length scale ($\lambda/2$) relative to the decay length of the low-temperature attractive force ($\lambda$):
\begin{equation}
f(x)\simeq \frac{t(x)}{T}t'(x)\propto e^{-2x/\lambda};\quad T\gg t(x),\,x\gg\lambda.
\end{equation}
Although it is suppressed by temperature, the above force may still overcome the magnetic repulsion at large separations provided $\xi<\lambda/2$.
For $\xi \ll \lambda/2$, we expect to see a free-energy minimum at $T>T^*$ which becomes shallow as the temperature is increased.
For $\lambda>\xi\gtrsim\lambda/2$,  we expect the minimum to vanish above $T^*$.
The two types of behavior can be seen in  Fig.~\ref{fig_finite_T}.

\section{Potential realization}
\label{sec_considerations}

A promising candidate material to host skyrmions that may realize this effect is the topological insulator heterostructure
$\textrm{Cr}_{x}{(\textrm{Bi}_{1-y}\textrm{Sb}_{y})}_{2-x}\textrm{Te}_{3} /
{(\textrm{Bi}_{1-y}\textrm{Sb}_{y})}_{2}\textrm{Te}_{3}$.
In
Ref.~[\onlinecite{2016_Yasuda}], Yasuda, \emph{et al.}~argue that this
heterostructure hosts a skyrmion ground state for  $\tilde{B}\lesssim
0.1$~T at $T\lesssim10$ K.
Band structure calculations suggest that this
material has an electronic gap of $\Delta\approx10$~meV and a Fermi velocity of $\hbar
v\approx500$~meV$\cdot$nm.\cite{2016_Yasuda} This suggests that skyrmions of
radius $R\gtrsim 50\,\mathrm{nm}$ would host skyrmion bound states in this material.
Accurate characterization of the actual skyrmion size, e.g., by magnetic atomic
force microscopy, would be necessary to ensure that this condition is fulfilled. 
For a tunnel coupling $t\qty(2R)=1\,\mathrm{meV}$, and skyrmion-skyrmion binding energy $\simeq 0.1\,\mathrm{meV}$, skyrmion-skyrmion bound states should form when $T\lesssim 1$~K.

Throughout this paper, we have neglected the repulsive dipolar interaction between skyrmions as well as the magnetic vector potential.  We have also assumed that double-occupancy of the molecular orbitals is suppressed by a weakly-screened Coulomb interaction. 
Here we argue that these approximations are justified for the heterostructure considered in Ref.~[\onlinecite{2016_Yasuda}] for skyrmions of radius $R=\lO=50$ nm  and with $t\qty(2R)=1$~meV. Such skyrmions will have $E_j/\Delta\approx-0.6$ for their lowest-energy single-skyrmion orbitals (see Fig.~\ref{fig_1sk_spectrum}), and will therefore have an orbital decay length, defined through Eq.~\eqref{eq_lambdadefn}, $\l\approx 60$~nm.

The repulsive dipolar interaction between skyrmions will dominate both the short-range repulsive interaction considered in Sec.~\ref{sec_msystem} and the short-range attractive interaction examined in Sec.~\ref{sec_esystem} at large separations. 
We approximate the dipolar interaction energy as $E_D \approx \frac{\mu_0\abs{\d}^2}{4\pi x^3}$ where $\d\approx\frac{\pi \alpha gs R^2 l}{V}  \mu_B \zh$ is the dipole moment due to the core of a skyrmion. 
We estimate $\alpha\approx0.6$ as the number of dopants per unit cell,\cite{2016_Yasuda} $V\approx0.59$~nm$^3$ as the unit cell volume,\cite{1972_Jenkins} $l\approx2$~nm as the depth of the magnetic layer,\cite{2016_Yasuda} and $g s\approx 1$ as the magnitude of the moment of the dopant atoms.
The force due to the dipolar interaction balances the force due to the attractive interaction at $x_D$ satisfying $\Phi'\qty(x_D) = E_D'\qty(x_D)$, with $\Phi\qty(x)$ defined in Eq.~\eqref{eq_GPdefn}.
Under the above assumptions, this equality is satisfied for  $x_D\approx450$~nm at $T=20$~mK. 
Below this separation, the repulsive dipolar force may be safely neglected.

In Eq.~\eqref{eq_MTIHamiltonian}, we have neglected the contribution of the magnetic vector potential by arguing that the $U(1)$ phase acquired by an electron in a skyrmion-bound orbital of radius $R=50$~nm is small for $\tilde{B}\ll 1$~T. This is easily satisfied for $\tilde{B}\lesssim 0.1$~T, the field strength at which the skyrmion crystal reported in Ref.~[\onlinecite{2016_Yasuda}] is stable. 

Finally, we assumed that the molecular orbitals support only $N=0$ or $1$ electrons because of a large charging energy. 
The doubly-occupied state will be irrelevant if the chemical potential is well below the sum of the excited molecular orbital energy and the charging energy, i.e.,~if $E_+\qty(x)+U\qty(x)-\mu\gg T$, where $U\qty(x)$ is the charging energy.
This is naturally satisfied at all separations if $E_0 - \mu \gg T$. 
Thus, the majority of the phase diagram, Fig.~\ref{fig_phasediagram}(a), is valid even for a vanishing Coulomb interaction, provided that the temperature is small compared to the binding energy.
For a weakly-screened Coulomb interaction, $U\qty(x) = \frac{e^2}{4\pi\epsilon}\frac{\exp(-\flatfrac{x}{\lD})}{ x}$, where $\lD$ is the Debye length and $\epsilon$ is the permittivity, the doubly-occupied state may be neglected for separations smaller than $x_C$ satisfying $E_+\qty(x_C)+U\qty(x_C)-\mu=T$. 
Assuming $\mu =E_0 $, $\lD=500$~nm, $\epsilon=\epsilon_0$ (i.e., $\frac{e^2}{4\pi \epsilon} = 1\,e$V-nm), and $T=20$~mK, we find $x_C\approx2.5$~$\mu$m.

Based on the length scales estimated above, our analysis will be applicable provided that the skyrmions are confined to a sample of size $L<\mathrm{min}\left(x_C, x_D\right)\approx 450$~nm. 
Larger samples may still host stable bound states, but the neglected effects will dominate the skyrmion-skyrmion interaction at large separations.

In addition to the promising MTI candidate material studied in Ref.~\onlinecite{2016_Yasuda}, there may also be other systems where it is possible to realize this effect, e.g.~a TI device functionalized with a magnetic top layer,\cite{2017_Gong, 2017_Huang} or even in a 2-dimensional topological insulator (2DTI) setting.  This can be realized in a 2DTI where there are two massive Dirac cones of unequal gap sizes (such as the Haldane model\cite{Haldane} with sublattice asymmetry\cite{Semenoff}).  If the coupling to the magnetic system changes the mass of the two valleys in the same way, there would be a regime in which a sign change in the magnetization amounts to a sign change in the Dirac mass of only one valley.  In the presence of skyrmions this would lead to electronic bound states of the kind discussed here.  

The DMI assumed here is essential to stabilize skyrmions.  This interaction is allowed in the bulk only for non-centrosymmetric materials.  However, an interface may also break inversion symmetry leading to a robust DMI (see, e.g., Ref.~\onlinecite{2016_Soumyanarayanan} for a review).  A magnetic thin film on top of a topological insulator may then naturally lead to the required DMI at the interface,\cite{2017_Zarzuela,2018_Zarzuela} without the need to specialize to non-centrosymmetric materials.

\section{Summary and conclusion}
In summary, we have shown that a pair of skyrmions on the surface of a magnetic
topological insulator may experience a mutually attractive interaction. In the
absence of the topological insulator's Dirac surface electrons, the skyrmions
will experience a repulsive interaction. The contribution of the electronic
system may be switched on and off by tuning the chemical potential. For large
chemical potential, the skyrmions form a bound state, while for low chemical
potential they remain unbound. For intermediate chemical potential, both the
bound and unbound states are locally stable. A chemical-potential sweep from low
to high and back should therefore lead to hysteretic binding and unbinding of
the skyrmion pair.
This hysteretic binding may find use in skyrmionic devices,\cite{2017_Fert} either as means of information storage, or as means to couple electronic and skyrmionic degrees of freedom.   
It may also be possible to generate GHz spin waves without microwave-frequency magnetic fields via an AC gate voltage. 
Such a magnon source could allow direct coupling of conventional and spin-wave circuits.  
The conditions laid out here for skyrmion binding could furthermore be used to realize stable qubits from the resulting molecular states.\cite{2013_Ferreira}
Beyond the predictions made here, the hybridization of skyrmion-bound orbitals may lead to novel magnetoelectric effects within the skyrmion crystal phase, and the contribution of the electronic free energy may modify the skyrmion crystal phase boundary.   

There is significant experimental effort in growing and characterizing magnetic topological insulators capable of supporting quantized anomalous Hall conductance and related domain-wall-bound-state phenomena.
Current state-of-the-art Cr doped
    ${\qty(\mathrm{Bi}_{2-y}\mathrm{Sb}_{y})}_{2}\mathrm{Te}_3$ heterostructures show signs of a skyrmion crystal phase,\cite{2016_Yasuda,2017_Liu,2019_Jiang} while the recent discovery of easy-axis ferromagnetic van der Waals materials may lead to a new class of magnetic topological insulators.\cite{2017_Gong,2017_Huang}
Magnetic topological insulators proximity-coupled to a ferromagnetic thin film have already been shown to host skyrmions.\cite{2018_Zhang}  Observing the effect considered in this work may be a natural next step toward demonstrating that
skyrmion physics on the surface of a topological insulator leads to new and
exciting effects.

\begin{acknowledgments}
    This work was enabled in part by support provided by NSERC, FRQNT, INTRIQ, CIFAR, Nordea Fonden, the FRQNT doctoral scholarship, and Compute Canada. 
\end{acknowledgments}

\appendix

\section{Numerical solution of Eq.~\eqref{eq_FMfdv}}
\label{ap_2sknumerics}
In the main text, we introduce a phenomenological form of $\FM\qty(x)$, the magnetic free
energy for two skyrmions separated by distance $x$.
Since we have no exact analytical solution for the free energy, $\FM(x)$, or generally
Eq.~\eqref{eq_FMfdv} with arbitrary boundary conditions, we must numerically
determine $\m\qty(\r)$ and $\FM\qty(x)$ to justify our phenomenological model.
We outline our numerical approach to minimize the magnetic free energy here. 

To minimize $\FM$, we evolve $\m$ with the partial differential equation
\begin{align}
    \pdv{\m}{\tau} = \vb{H}-\qty(\m\cdot\vb{H})\m\label{eq_relax}
\end{align}
where $\tau$ is the simulation time and 
\begin{align}
    \vb{H} = -\fdv{F_M}{\m}.
\end{align}
Eq.~\eqref{eq_relax} is simply the component of the Landau-Lifschitz-Gilbert
equation\cite{2004_Gilbert} that leads to relaxation.
In the long-time limit,
when $\pdv{\m}{\tau}=0$, the magnetization has been evolved to a configuration
that minimizes $\FM\qty[\m]$ under the constraint $\abs{\m}=1$. 
We implement boundary conditions by setting
$\vb{H}\qty(\r)=0$ for $\r$ at boundaries and choosing the correct initial
condition. 

To find $\FM\qty(x)$, we initialize the magnetization in an approximate
configuration in the two-skyrmion topological sector that should be close to the
ground state. We separate the magnetization into a left and a right region and
use the numerical solution of Eq.~\eqref{eq_1skODE} to prepare a single skyrmion
centered at $-\flatfrac{x'}{2}$ in the left region. Similarly, we prepare a
skyrmion at $\flatfrac{x'}{2}$ in the right region.  Then we fix $\vb{H}\qty(\pm
\flatfrac{x'\xh}{2})=0$ to pin the skyrmion cores and evolve the magnetization
under Eq.~\eqref{eq_relax} until dynamics cease. From the resulting
magnetization we can numerically evaluate Eq.~\eqref{eq_FMtotal} to find $\FM$.
However, the pinned skyrmion-skyrmion separation, $x'$, is not the distance
between the skyrmion centers. Since the skyrmions repel, they will move away
from each other under Eq.~\eqref{eq_relax} until the pinned magnetic moments
reach the skyrmion perimeters. We determine the location of the skyrmion
centers, $\r_\pm$, by fitting the final $m_z$ with the ansatz 
\begin{align}
    \Theta\qty(\r) =& 2\sum_{i\in\pm}\arctan\qty(\frac{\sinh{\flatfrac{R}{\xi
    }}}{\sinh{\flatfrac{\rho_i}{\xi}}})\\ 
    {\rho}_{\pm} =& \abs{\r-\r_\pm},
\end{align}
where $m_z = \cos\Theta$. This fitting procedure is justified in the limit
$\flatfrac{\qty(x-2R)}{\xi} \gg 1$.

\section{Numerical determination of single-skyrmion instability phase boundary}
\label{ap_stability}
In the Sec.~\ref{sec_msystem}, we argue that the magnetic free energy is stationary for a skyrmion texture, i.e. Eq.~\eqref{eq_FMfdv} is solved by Eq.~\eqref{eq_1skM}, with $\Theta$ given by the solution of Eq.~\eqref{eq_1skODE}.  
While this is true, for sufficiently low $\qty(B,K)$, the single-skyrmion texture does not minimize $\FM$. 
It is instead a saddle-point solution. 
In this region of parameter space, skyrmions are unstable to transitions towards a spin-spiral phase. 
This phase boundary is well known in the literature,\cite{2014_Banerjee} and is identified by our numerical simulations. 
For $\qty(B,K)$ below the instability phase boundary, we find that
Eq.~\eqref{eq_relax}, which minimizes $F_M$ under the constraint $\abs{\m}=1$, takes an initial single-skyrmion configurations to 
a final spiral configurations.  
To determine the phase boundary, we determine
the lifetime of the skyrmion configuration, prepared using the numerical solution of Eq.~\eqref{eq_1skODE}, under evolution with
Eq.~\eqref{eq_relax}. We define the lifetime to be the time elapsed before the
magnetization deviates from cylindrical symmetry. Specifically, we calculate the
time $\taus$ at which the $z$ component of the magnetization deviates from its
azimuthal average by a small threshold:
\begin{align}
\int \dd{\r}{\qty(m_z\qty(\r,\taus) - \ev{m_z\qty(\r,\taus)}_\phi)}^{2} =
\delta
\end{align}
where 
\begin{align}
\ev{m_z}_\phi = \frac{1}{2\pi}\int \dd{\phi} m_z\qty(r,\phi).
\end{align}
The skyrmion lifetime diverges at the instability phase boundary. We fit
$\taus\qty(B,K)$ for $B,K$ within the unstable phase using 
\begin{align}
\taus\qty(B,K) = {A\qty(K-\Ks\qty(B))}^\alpha
\end{align}
to determine $\Ks\qty(B)$, the critical anisotropy for a given applied field.
The phase boundary plotted in Fig.~\ref{fig_phasediagram} is determined by these
$\Ks\qty(B)$ for varying $B$. This phase boundary is consistent with the
phase diagram presented in Ref.~[\onlinecite{2014_Banerjee}]. 

\section{Numerical calculation of in-gap electronic states}\label{ap_electronic}
To justify our tight-binding model and verify the analytical single-skyrmion
orbitals, we determine the electronic surface states numerically on a lattice.
We can construct a lattice model whose low-energy excitations are governed by
Eq.~\eqref{eq_MTIHamiltonian}.  Following Ref.~[\onlinecite{2012_Marchand}], we
consider an electronic system governed by \begin{align} H = \sum_{\k}\ckd
h_{\k}\ck + \Delta\sum_i \psi_i\bsigma\cdot\m_i\psi_i, \end{align} where
\begin{align} 
    h_{\k} =& v\qty[\sin(k_x)\sigma_y - \sin(k_y)\sigma_x]
    \nonumber\\&+
    \eta\qty[2-\cos(k_x)-\cos(k_y)]\sigma_z.\label{eq_numerical_bloch}
\end{align} 
The term proportional to $\eta$ in Eq.~\eqref{eq_numerical_bloch} gaps the
lattice model's Dirac cones at the edge of the Brillouin zone, leaving
low-energy excitations only around $\k = 0$. Thus for $\eta >\Delta$, the
low-energy, long-wavelength, physics of this model should be given by
Eq.~\eqref{eq_MTIHamiltonian}. 

We used the python package Kwant\cite{2014_Groth} to determine the electronic
structure associated with the magnetization generated from the simulations described
in Appendix \ref{ap_2sknumerics}.

\bibliography{bib}

\begin{thebibliography}{41}
\expandafter\ifx\csname natexlab\endcsname\relax\def\natexlab#1{#1}\fi
\expandafter\ifx\csname bibnamefont\endcsname\relax
  \def\bibnamefont#1{#1}\fi
\expandafter\ifx\csname bibfnamefont\endcsname\relax
  \def\bibfnamefont#1{#1}\fi
\expandafter\ifx\csname citenamefont\endcsname\relax
  \def\citenamefont#1{#1}\fi
\expandafter\ifx\csname url\endcsname\relax
  \def\url#1{\texttt{#1}}\fi
\expandafter\ifx\csname urlprefix\endcsname\relax\def\urlprefix{URL }\fi
\providecommand{\bibinfo}[2]{#2}
\providecommand{\eprint}[2][]{\url{#2}}

\bibitem[{\citenamefont{Hasan and Kane}(2010)}]{2010_Hasan}
\bibinfo{author}{\bibfnamefont{M.~Z.} \bibnamefont{Hasan}} \bibnamefont{and}
  \bibinfo{author}{\bibfnamefont{C.~L.} \bibnamefont{Kane}},
  \bibinfo{journal}{Rev. Mod. Phys.} \textbf{\bibinfo{volume}{82}},
  \bibinfo{pages}{3045} (\bibinfo{year}{2010}),
  \urlprefix\url{https://link.aps.org/doi/10.1103/RevModPhys.82.3045}.

\bibitem[{\citenamefont{Bansil et~al.}(2016)\citenamefont{Bansil, Lin, and
  Das}}]{2016_Bansil}
\bibinfo{author}{\bibfnamefont{A.}~\bibnamefont{Bansil}},
  \bibinfo{author}{\bibfnamefont{H.}~\bibnamefont{Lin}}, \bibnamefont{and}
  \bibinfo{author}{\bibfnamefont{T.}~\bibnamefont{Das}}, \bibinfo{journal}{Rev.
  Mod. Phys.} \textbf{\bibinfo{volume}{88}}, \bibinfo{pages}{021004}
  (\bibinfo{year}{2016}),
  \urlprefix\url{https://link.aps.org/doi/10.1103/RevModPhys.88.021004}.

\bibitem[{\citenamefont{Chiu et~al.}(2016)\citenamefont{Chiu, Teo, Schnyder,
  and Ryu}}]{2016_Chiu}
\bibinfo{author}{\bibfnamefont{C.-K.} \bibnamefont{Chiu}},
  \bibinfo{author}{\bibfnamefont{J.~C.~Y.} \bibnamefont{Teo}},
  \bibinfo{author}{\bibfnamefont{A.~P.} \bibnamefont{Schnyder}},
  \bibnamefont{and} \bibinfo{author}{\bibfnamefont{S.}~\bibnamefont{Ryu}},
  \bibinfo{journal}{Rev. Mod. Phys.} \textbf{\bibinfo{volume}{88}},
  \bibinfo{pages}{035005} (\bibinfo{year}{2016}),
  \urlprefix\url{https://link.aps.org/doi/10.1103/RevModPhys.88.035005}.

\bibitem[{\citenamefont{Liu et~al.}(2009)\citenamefont{Liu, Liu, Xu, Qi, and
  Zhang}}]{2009_Liu}
\bibinfo{author}{\bibfnamefont{Q.}~\bibnamefont{Liu}},
  \bibinfo{author}{\bibfnamefont{C.-X.} \bibnamefont{Liu}},
  \bibinfo{author}{\bibfnamefont{C.}~\bibnamefont{Xu}},
  \bibinfo{author}{\bibfnamefont{X.-L.} \bibnamefont{Qi}}, \bibnamefont{and}
  \bibinfo{author}{\bibfnamefont{S.-C.} \bibnamefont{Zhang}},
  \bibinfo{journal}{Phys. Rev. Lett.} \textbf{\bibinfo{volume}{102}},
  \bibinfo{pages}{156603} (\bibinfo{year}{2009}),
  \urlprefix\url{https://link.aps.org/doi/10.1103/PhysRevLett.102.156603}.

\bibitem[{\citenamefont{Zhu et~al.}(2011)\citenamefont{Zhu, Yao, Zhang, and
  Chang}}]{2011_Zhu}
\bibinfo{author}{\bibfnamefont{J.-J.} \bibnamefont{Zhu}},
  \bibinfo{author}{\bibfnamefont{D.-X.} \bibnamefont{Yao}},
  \bibinfo{author}{\bibfnamefont{S.-C.} \bibnamefont{Zhang}}, \bibnamefont{and}
  \bibinfo{author}{\bibfnamefont{K.}~\bibnamefont{Chang}},
  \bibinfo{journal}{Phys. Rev. Lett.} \textbf{\bibinfo{volume}{106}},
  \bibinfo{pages}{097201} (\bibinfo{year}{2011}),
  \urlprefix\url{https://link.aps.org/doi/10.1103/PhysRevLett.106.097201}.

\bibitem[{\citenamefont{Tokura et~al.}(2019)\citenamefont{Tokura, Yasuda, and
  Tsukazaki}}]{2019_Tokura}
\bibinfo{author}{\bibfnamefont{Y.}~\bibnamefont{Tokura}},
  \bibinfo{author}{\bibfnamefont{K.}~\bibnamefont{Yasuda}}, \bibnamefont{and}
  \bibinfo{author}{\bibfnamefont{A.}~\bibnamefont{Tsukazaki}},
  \bibinfo{journal}{Nature Reviews Physics} \textbf{\bibinfo{volume}{1}},
  \bibinfo{pages}{126} (\bibinfo{year}{2019}), ISSN \bibinfo{issn}{2522-5820},
  \urlprefix\url{https://doi.org/10.1038/s42254-018-0011-5}.

\bibitem[{\citenamefont{Jackiw and Rebbi}(1976)}]{1976_Jackiw}
\bibinfo{author}{\bibfnamefont{R.}~\bibnamefont{Jackiw}} \bibnamefont{and}
  \bibinfo{author}{\bibfnamefont{C.}~\bibnamefont{Rebbi}},
  \bibinfo{journal}{Phys. Rev. D} \textbf{\bibinfo{volume}{13}},
  \bibinfo{pages}{3398} (\bibinfo{year}{1976}),
  \urlprefix\url{https://link.aps.org/doi/10.1103/PhysRevD.13.3398}.

\bibitem[{\citenamefont{Jackiw}(1984)}]{1984_Jackiw}
\bibinfo{author}{\bibfnamefont{R.}~\bibnamefont{Jackiw}},
  \bibinfo{journal}{Phys. Rev. D} \textbf{\bibinfo{volume}{29}},
  \bibinfo{pages}{2375} (\bibinfo{year}{1984}),
  \urlprefix\url{https://link.aps.org/doi/10.1103/PhysRevD.29.2375}.

\bibitem[{\citenamefont{Chang et~al.}(2013)\citenamefont{Chang, Zhang, Feng,
  Shen, Zhang, Guo, Li, Ou, Wei, Wang et~al.}}]{2013_Chang}
\bibinfo{author}{\bibfnamefont{C.-Z.} \bibnamefont{Chang}},
  \bibinfo{author}{\bibfnamefont{J.}~\bibnamefont{Zhang}},
  \bibinfo{author}{\bibfnamefont{X.}~\bibnamefont{Feng}},
  \bibinfo{author}{\bibfnamefont{J.}~\bibnamefont{Shen}},
  \bibinfo{author}{\bibfnamefont{Z.}~\bibnamefont{Zhang}},
  \bibinfo{author}{\bibfnamefont{M.}~\bibnamefont{Guo}},
  \bibinfo{author}{\bibfnamefont{K.}~\bibnamefont{Li}},
  \bibinfo{author}{\bibfnamefont{Y.}~\bibnamefont{Ou}},
  \bibinfo{author}{\bibfnamefont{P.}~\bibnamefont{Wei}},
  \bibinfo{author}{\bibfnamefont{L.-L.} \bibnamefont{Wang}},
  \bibnamefont{et~al.}, \bibinfo{journal}{Science}
  \textbf{\bibinfo{volume}{340}}, \bibinfo{pages}{167} (\bibinfo{year}{2013}),
  ISSN \bibinfo{issn}{0036-8075},
  \urlprefix\url{http://science.sciencemag.org/content/340/6129/167}.

\bibitem[{\citenamefont{Nakajima et~al.}(2015)\citenamefont{Nakajima, Syers,
  Wang, Wang, and Paglione}}]{2015_Nakajima}
\bibinfo{author}{\bibfnamefont{Y.}~\bibnamefont{Nakajima}},
  \bibinfo{author}{\bibfnamefont{P.}~\bibnamefont{Syers}},
  \bibinfo{author}{\bibfnamefont{X.}~\bibnamefont{Wang}},
  \bibinfo{author}{\bibfnamefont{R.}~\bibnamefont{Wang}}, \bibnamefont{and}
  \bibinfo{author}{\bibfnamefont{J.}~\bibnamefont{Paglione}},
  \bibinfo{journal}{Nature Physics} \textbf{\bibinfo{volume}{12}},
  \bibinfo{pages}{213 EP } (\bibinfo{year}{2015}),
  \urlprefix\url{https://doi.org/10.1038/nphys3555}.

\bibitem[{\citenamefont{Tiwari et~al.}(2017)\citenamefont{Tiwari, Coish, and
  Pereg-Barnea}}]{2017_Tiwari}
\bibinfo{author}{\bibfnamefont{K.~L.} \bibnamefont{Tiwari}},
  \bibinfo{author}{\bibfnamefont{W.~A.} \bibnamefont{Coish}}, \bibnamefont{and}
  \bibinfo{author}{\bibfnamefont{T.}~\bibnamefont{Pereg-Barnea}},
  \bibinfo{journal}{Phys. Rev. B} \textbf{\bibinfo{volume}{96}},
  \bibinfo{pages}{235120} (\bibinfo{year}{2017}),
  \urlprefix\url{https://link.aps.org/doi/10.1103/PhysRevB.96.235120}.

\bibitem[{\citenamefont{Hurst et~al.}(2015)\citenamefont{Hurst, Efimkin, Zang,
  and Galitski}}]{2015_Hurst}
\bibinfo{author}{\bibfnamefont{H.~M.} \bibnamefont{Hurst}},
  \bibinfo{author}{\bibfnamefont{D.~K.} \bibnamefont{Efimkin}},
  \bibinfo{author}{\bibfnamefont{J.}~\bibnamefont{Zang}}, \bibnamefont{and}
  \bibinfo{author}{\bibfnamefont{V.}~\bibnamefont{Galitski}},
  \bibinfo{journal}{Phys. Rev. B} \textbf{\bibinfo{volume}{91}},
  \bibinfo{pages}{060401(R)} (\bibinfo{year}{2015}),
  \urlprefix\url{https://link.aps.org/doi/10.1103/PhysRevB.91.060401}.

\bibitem[{\citenamefont{Nagaosa and Tokura}(2013)}]{2013_Nagaosa}
\bibinfo{author}{\bibfnamefont{N.}~\bibnamefont{Nagaosa}} \bibnamefont{and}
  \bibinfo{author}{\bibfnamefont{Y.}~\bibnamefont{Tokura}},
  \bibinfo{journal}{Nature Nanotechnology} \textbf{\bibinfo{volume}{8}},
  \bibinfo{pages}{899 EP } (\bibinfo{year}{2013}),
  \urlprefix\url{https://doi.org/10.1038/nnano.2013.243}.

\bibitem[{\citenamefont{R{\"o}{\ss}ler
  et~al.}(2006)\citenamefont{R{\"o}{\ss}ler, Bogdanov, and
  Pfleiderer}}]{2006_Rossler}
\bibinfo{author}{\bibfnamefont{U.~K.} \bibnamefont{R{\"o}{\ss}ler}},
  \bibinfo{author}{\bibfnamefont{A.~N.} \bibnamefont{Bogdanov}},
  \bibnamefont{and}
  \bibinfo{author}{\bibfnamefont{C.}~\bibnamefont{Pfleiderer}},
  \bibinfo{journal}{Nature} \textbf{\bibinfo{volume}{442}}, \bibinfo{pages}{797
  EP } (\bibinfo{year}{2006}),
  \urlprefix\url{https://doi.org/10.1038/nature05056}.

\bibitem[{\citenamefont{Ferreira and Loss}(2013)}]{2013_Ferreira}
\bibinfo{author}{\bibfnamefont{G.~J.} \bibnamefont{Ferreira}} \bibnamefont{and}
  \bibinfo{author}{\bibfnamefont{D.}~\bibnamefont{Loss}},
  \bibinfo{journal}{\prl} \textbf{\bibinfo{volume}{111}},
  \bibinfo{pages}{106802} (\bibinfo{year}{2013}).

\bibitem[{\citenamefont{Uchoa et~al.}(2015)\citenamefont{Uchoa, Kotov, and
  Kindermann}}]{2015_Uchoa}
\bibinfo{author}{\bibfnamefont{B.}~\bibnamefont{Uchoa}},
  \bibinfo{author}{\bibfnamefont{V.~N.} \bibnamefont{Kotov}}, \bibnamefont{and}
  \bibinfo{author}{\bibfnamefont{M.}~\bibnamefont{Kindermann}},
  \bibinfo{journal}{\prb} \textbf{\bibinfo{volume}{91}},
  \bibinfo{pages}{121412(R)} (\bibinfo{year}{2015}).

\bibitem[{\citenamefont{Lin et~al.}(2013)\citenamefont{Lin, Reichhardt,
  Batista, and Saxena}}]{2013_Lin}
\bibinfo{author}{\bibfnamefont{S.-Z.} \bibnamefont{Lin}},
  \bibinfo{author}{\bibfnamefont{C.}~\bibnamefont{Reichhardt}},
  \bibinfo{author}{\bibfnamefont{C.~D.} \bibnamefont{Batista}},
  \bibnamefont{and} \bibinfo{author}{\bibfnamefont{A.}~\bibnamefont{Saxena}},
  \bibinfo{journal}{Phys. Rev. B} \textbf{\bibinfo{volume}{87}},
  \bibinfo{pages}{214419} (\bibinfo{year}{2013}),
  \urlprefix\url{https://link.aps.org/doi/10.1103/PhysRevB.87.214419}.

\bibitem[{\citenamefont{Leonov and Mostovoy}(2015)}]{2015_Leonov}
\bibinfo{author}{\bibfnamefont{A.~O.} \bibnamefont{Leonov}} \bibnamefont{and}
  \bibinfo{author}{\bibfnamefont{M.}~\bibnamefont{Mostovoy}},
  \bibinfo{journal}{Nature communications} \textbf{\bibinfo{volume}{6}},
  \bibinfo{pages}{8275} (\bibinfo{year}{2015}).

\bibitem[{\citenamefont{Kharkov et~al.}(2017)\citenamefont{Kharkov, Sushkov,
  and Mostovoy}}]{2017_Kharkov}
\bibinfo{author}{\bibfnamefont{Y.~A.} \bibnamefont{Kharkov}},
  \bibinfo{author}{\bibfnamefont{O.~P.} \bibnamefont{Sushkov}},
  \bibnamefont{and} \bibinfo{author}{\bibfnamefont{M.}~\bibnamefont{Mostovoy}},
  \bibinfo{journal}{\prl} \textbf{\bibinfo{volume}{119}},
  \bibinfo{pages}{207201} (\bibinfo{year}{2017}).

\bibitem[{\citenamefont{Fert et~al.}(2017)\citenamefont{Fert, Reyren, and
  Cros}}]{2017_Fert}
\bibinfo{author}{\bibfnamefont{A.}~\bibnamefont{Fert}},
  \bibinfo{author}{\bibfnamefont{N.}~\bibnamefont{Reyren}}, \bibnamefont{and}
  \bibinfo{author}{\bibfnamefont{V.}~\bibnamefont{Cros}},
  \bibinfo{journal}{Nat. Rev. Mater.} \textbf{\bibinfo{volume}{2}},
  \bibinfo{pages}{17031} (\bibinfo{year}{2017}).

\bibitem[{\citenamefont{Dzyaloshinsky}(1958)}]{1958_Dzyaloshinsky}
\bibinfo{author}{\bibfnamefont{I.}~\bibnamefont{Dzyaloshinsky}},
  \bibinfo{journal}{Journal of Physics and Chemistry of Solids}
  \textbf{\bibinfo{volume}{4}}, \bibinfo{pages}{241 } (\bibinfo{year}{1958}),
  ISSN \bibinfo{issn}{0022-3697},
  \urlprefix\url{http://www.sciencedirect.com/science/article/pii/0022369758900763}.

\bibitem[{\citenamefont{Moriya}(1960)}]{1960_Moriya}
\bibinfo{author}{\bibfnamefont{T.}~\bibnamefont{Moriya}},
  \bibinfo{journal}{Phys. Rev.} \textbf{\bibinfo{volume}{120}},
  \bibinfo{pages}{91} (\bibinfo{year}{1960}),
  \urlprefix\url{https://link.aps.org/doi/10.1103/PhysRev.120.91}.

\bibitem[{\citenamefont{G\"ung\"ord\"u
  et~al.}(2016)\citenamefont{G\"ung\"ord\"u, Nepal, Tretiakov, Belashchenko,
  and Kovalev}}]{2016_Utkan}
\bibinfo{author}{\bibfnamefont{U.}~\bibnamefont{G\"ung\"ord\"u}},
  \bibinfo{author}{\bibfnamefont{R.}~\bibnamefont{Nepal}},
  \bibinfo{author}{\bibfnamefont{O.~A.} \bibnamefont{Tretiakov}},
  \bibinfo{author}{\bibfnamefont{K.}~\bibnamefont{Belashchenko}},
  \bibnamefont{and} \bibinfo{author}{\bibfnamefont{A.~A.}
  \bibnamefont{Kovalev}}, \bibinfo{journal}{Phys. Rev. B}
  \textbf{\bibinfo{volume}{93}}, \bibinfo{pages}{064428}
  (\bibinfo{year}{2016}),
  \urlprefix\url{https://link.aps.org/doi/10.1103/PhysRevB.93.064428}.

\bibitem[{\citenamefont{Banerjee et~al.}(2014)\citenamefont{Banerjee, Rowland,
  Erten, and Randeria}}]{2014_Banerjee}
\bibinfo{author}{\bibfnamefont{S.}~\bibnamefont{Banerjee}},
  \bibinfo{author}{\bibfnamefont{J.}~\bibnamefont{Rowland}},
  \bibinfo{author}{\bibfnamefont{O.}~\bibnamefont{Erten}}, \bibnamefont{and}
  \bibinfo{author}{\bibfnamefont{M.}~\bibnamefont{Randeria}},
  \bibinfo{journal}{Phys. Rev. X} \textbf{\bibinfo{volume}{4}},
  \bibinfo{pages}{031045} (\bibinfo{year}{2014}),
  \urlprefix\url{https://link.aps.org/doi/10.1103/PhysRevX.4.031045}.

\bibitem[{\citenamefont{Han}(2017)}]{2017_Han}
\bibinfo{author}{\bibfnamefont{J.~H.} \bibnamefont{Han}},
  \emph{\bibinfo{title}{Skyrmions in Condensed Matter}}, Springer tracts in
  modern physics, 0081-3869 ; volume 278 (\bibinfo{publisher}{Springer},
  \bibinfo{address}{Cham, Switzerland}, \bibinfo{year}{2017}).

\bibitem[{\citenamefont{Yasuda et~al.}(2016)\citenamefont{Yasuda, Wakatsuki,
  Morimoto, Yoshimi, Tsukazaki, Takahashi, Ezawa, Kawasaki, Nagaosa, and
  Tokura}}]{2016_Yasuda}
\bibinfo{author}{\bibfnamefont{K.}~\bibnamefont{Yasuda}},
  \bibinfo{author}{\bibfnamefont{R.}~\bibnamefont{Wakatsuki}},
  \bibinfo{author}{\bibfnamefont{T.}~\bibnamefont{Morimoto}},
  \bibinfo{author}{\bibfnamefont{R.}~\bibnamefont{Yoshimi}},
  \bibinfo{author}{\bibfnamefont{A.}~\bibnamefont{Tsukazaki}},
  \bibinfo{author}{\bibfnamefont{K.~S.} \bibnamefont{Takahashi}},
  \bibinfo{author}{\bibfnamefont{M.}~\bibnamefont{Ezawa}},
  \bibinfo{author}{\bibfnamefont{M.}~\bibnamefont{Kawasaki}},
  \bibinfo{author}{\bibfnamefont{N.}~\bibnamefont{Nagaosa}}, \bibnamefont{and}
  \bibinfo{author}{\bibfnamefont{Y.}~\bibnamefont{Tokura}},
  \bibinfo{journal}{Nature Physics} \textbf{\bibinfo{volume}{12}},
  \bibinfo{pages}{555 EP } (\bibinfo{year}{2016}),
  \urlprefix\url{https://doi.org/10.1038/nphys3671}.

\bibitem[{\citenamefont{Jenkins et~al.}(1972)\citenamefont{Jenkins, Rayne, and
  Ure}}]{1972_Jenkins}
\bibinfo{author}{\bibfnamefont{J.~O.} \bibnamefont{Jenkins}},
  \bibinfo{author}{\bibfnamefont{J.~A.} \bibnamefont{Rayne}}, \bibnamefont{and}
  \bibinfo{author}{\bibfnamefont{R.~W.} \bibnamefont{Ure}},
  \bibinfo{journal}{Phys. Rev. B} \textbf{\bibinfo{volume}{5}},
  \bibinfo{pages}{3171} (\bibinfo{year}{1972}),
  \urlprefix\url{https://link.aps.org/doi/10.1103/PhysRevB.5.3171}.

\bibitem[{\citenamefont{Gong et~al.}(2017)\citenamefont{Gong, Li, Li, Ji,
  Stern, Xia, Cao, Bao, Wang, Wang et~al.}}]{2017_Gong}
\bibinfo{author}{\bibfnamefont{C.}~\bibnamefont{Gong}},
  \bibinfo{author}{\bibfnamefont{L.}~\bibnamefont{Li}},
  \bibinfo{author}{\bibfnamefont{Z.}~\bibnamefont{Li}},
  \bibinfo{author}{\bibfnamefont{H.}~\bibnamefont{Ji}},
  \bibinfo{author}{\bibfnamefont{A.}~\bibnamefont{Stern}},
  \bibinfo{author}{\bibfnamefont{Y.}~\bibnamefont{Xia}},
  \bibinfo{author}{\bibfnamefont{T.}~\bibnamefont{Cao}},
  \bibinfo{author}{\bibfnamefont{W.}~\bibnamefont{Bao}},
  \bibinfo{author}{\bibfnamefont{C.}~\bibnamefont{Wang}},
  \bibinfo{author}{\bibfnamefont{Y.}~\bibnamefont{Wang}}, \bibnamefont{et~al.},
  \textbf{\bibinfo{volume}{546}}, \bibinfo{pages}{265 EP }
  (\bibinfo{year}{2017}), \urlprefix\url{https://doi.org/10.1038/nature22060}.

\bibitem[{\citenamefont{Huang et~al.}(2017)\citenamefont{Huang, Clark,
  Navarro-Moratalla, Klein, Cheng, Seyler, Zhong, Schmidgall, McGuire, Cobden
  et~al.}}]{2017_Huang}
\bibinfo{author}{\bibfnamefont{B.}~\bibnamefont{Huang}},
  \bibinfo{author}{\bibfnamefont{G.}~\bibnamefont{Clark}},
  \bibinfo{author}{\bibfnamefont{E.}~\bibnamefont{Navarro-Moratalla}},
  \bibinfo{author}{\bibfnamefont{D.~R.} \bibnamefont{Klein}},
  \bibinfo{author}{\bibfnamefont{R.}~\bibnamefont{Cheng}},
  \bibinfo{author}{\bibfnamefont{K.~L.} \bibnamefont{Seyler}},
  \bibinfo{author}{\bibfnamefont{D.}~\bibnamefont{Zhong}},
  \bibinfo{author}{\bibfnamefont{E.}~\bibnamefont{Schmidgall}},
  \bibinfo{author}{\bibfnamefont{M.~A.} \bibnamefont{McGuire}},
  \bibinfo{author}{\bibfnamefont{D.~H.} \bibnamefont{Cobden}},
  \bibnamefont{et~al.}, \bibinfo{journal}{Nature}
  \textbf{\bibinfo{volume}{546}}, \bibinfo{pages}{270 EP }
  (\bibinfo{year}{2017}), \urlprefix\url{https://doi.org/10.1038/nature22391}.

\bibitem[{\citenamefont{Haldane}(1988)}]{Haldane}
\bibinfo{author}{\bibfnamefont{F.~D.~M.} \bibnamefont{Haldane}},
  \bibinfo{journal}{Phys. Rev. Lett.} \textbf{\bibinfo{volume}{61}},
  \bibinfo{pages}{2015} (\bibinfo{year}{1988}),
  \urlprefix\url{https://link.aps.org/doi/10.1103/PhysRevLett.61.2015}.

\bibitem[{\citenamefont{Semenoff}(1984)}]{Semenoff}
\bibinfo{author}{\bibfnamefont{G.~W.} \bibnamefont{Semenoff}},
  \bibinfo{journal}{Phys. Rev. Lett.} \textbf{\bibinfo{volume}{53}},
  \bibinfo{pages}{2449} (\bibinfo{year}{1984}),
  \urlprefix\url{https://link.aps.org/doi/10.1103/PhysRevLett.53.2449}.

\bibitem[{\citenamefont{Soumyanarayanan
  et~al.}(2016)\citenamefont{Soumyanarayanan, Reyren, Fert, and
  Panagopoulos}}]{2016_Soumyanarayanan}
\bibinfo{author}{\bibfnamefont{A.}~\bibnamefont{Soumyanarayanan}},
  \bibinfo{author}{\bibfnamefont{N.}~\bibnamefont{Reyren}},
  \bibinfo{author}{\bibfnamefont{A.}~\bibnamefont{Fert}}, \bibnamefont{and}
  \bibinfo{author}{\bibfnamefont{C.}~\bibnamefont{Panagopoulos}},
  \bibinfo{journal}{Nature} \textbf{\bibinfo{volume}{539}},
  \bibinfo{pages}{509} (\bibinfo{year}{2016}).

\bibitem[{\citenamefont{Zarzuela and Tserkovnyak}(2017)}]{2017_Zarzuela}
\bibinfo{author}{\bibfnamefont{R.}~\bibnamefont{Zarzuela}} \bibnamefont{and}
  \bibinfo{author}{\bibfnamefont{Y.}~\bibnamefont{Tserkovnyak}},
  \bibinfo{journal}{\prb} \textbf{\bibinfo{volume}{95}},
  \bibinfo{pages}{180402(R)} (\bibinfo{year}{2017}).

\bibitem[{\citenamefont{Zarzuela et~al.}(2018)\citenamefont{Zarzuela, Kim, and
  Tserkovnyak}}]{2018_Zarzuela}
\bibinfo{author}{\bibfnamefont{R.}~\bibnamefont{Zarzuela}},
  \bibinfo{author}{\bibfnamefont{S.~K.} \bibnamefont{Kim}}, \bibnamefont{and}
  \bibinfo{author}{\bibfnamefont{Y.}~\bibnamefont{Tserkovnyak}},
  \bibinfo{journal}{\prb} \textbf{\bibinfo{volume}{97}},
  \bibinfo{pages}{014418} (\bibinfo{year}{2018}).

\bibitem[{\citenamefont{Liu et~al.}(2017)\citenamefont{Liu, Zang, Ruan, Gong,
  He, Ma, Xue, and Wang}}]{2017_Liu}
\bibinfo{author}{\bibfnamefont{C.}~\bibnamefont{Liu}},
  \bibinfo{author}{\bibfnamefont{Y.}~\bibnamefont{Zang}},
  \bibinfo{author}{\bibfnamefont{W.}~\bibnamefont{Ruan}},
  \bibinfo{author}{\bibfnamefont{Y.}~\bibnamefont{Gong}},
  \bibinfo{author}{\bibfnamefont{K.}~\bibnamefont{He}},
  \bibinfo{author}{\bibfnamefont{X.}~\bibnamefont{Ma}},
  \bibinfo{author}{\bibfnamefont{Q.-K.} \bibnamefont{Xue}}, \bibnamefont{and}
  \bibinfo{author}{\bibfnamefont{Y.}~\bibnamefont{Wang}},
  \bibinfo{journal}{Phys. Rev. Lett.} \textbf{\bibinfo{volume}{119}},
  \bibinfo{pages}{176809} (\bibinfo{year}{2017}),
  \urlprefix\url{https://link.aps.org/doi/10.1103/PhysRevLett.119.176809}.

\bibitem[{\citenamefont{Jiang et~al.}(2019)\citenamefont{Jiang, Xiao, Wang,
  Shin, Andreoli, Zhang, Xiao, Zhao, Kayyalha, Zhang et~al.}}]{2019_Jiang}
\bibinfo{author}{\bibfnamefont{J.}~\bibnamefont{Jiang}},
  \bibinfo{author}{\bibfnamefont{D.}~\bibnamefont{Xiao}},
  \bibinfo{author}{\bibfnamefont{F.}~\bibnamefont{Wang}},
  \bibinfo{author}{\bibfnamefont{J.-H.} \bibnamefont{Shin}},
  \bibinfo{author}{\bibfnamefont{D.}~\bibnamefont{Andreoli}},
  \bibinfo{author}{\bibfnamefont{J.}~\bibnamefont{Zhang}},
  \bibinfo{author}{\bibfnamefont{R.}~\bibnamefont{Xiao}},
  \bibinfo{author}{\bibfnamefont{Y.-F.} \bibnamefont{Zhao}},
  \bibinfo{author}{\bibfnamefont{M.}~\bibnamefont{Kayyalha}},
  \bibinfo{author}{\bibfnamefont{L.}~\bibnamefont{Zhang}},
  \bibnamefont{et~al.}, \bibinfo{journal}{arXiv:1901.07611}
  (\bibinfo{year}{2019}).

\bibitem[{\citenamefont{Zhang et~al.}(2018)\citenamefont{Zhang, Kronast,
  van~der Laan, and Hesjedal}}]{2018_Zhang}
\bibinfo{author}{\bibfnamefont{S.}~\bibnamefont{Zhang}},
  \bibinfo{author}{\bibfnamefont{F.}~\bibnamefont{Kronast}},
  \bibinfo{author}{\bibfnamefont{G.}~\bibnamefont{van~der Laan}},
  \bibnamefont{and} \bibinfo{author}{\bibfnamefont{T.}~\bibnamefont{Hesjedal}},
  \bibinfo{journal}{Nano letters} \textbf{\bibinfo{volume}{18}},
  \bibinfo{pages}{1057} (\bibinfo{year}{2018}).

\bibitem[{\citenamefont{Gilbert}(2004)}]{2004_Gilbert}
\bibinfo{author}{\bibfnamefont{T.}~\bibnamefont{Gilbert}},
  \bibinfo{journal}{IEEE Trans. Magn.} \textbf{\bibinfo{volume}{40}},
  \bibinfo{pages}{3443} (\bibinfo{year}{2004}).

\bibitem[{\citenamefont{Marchand and Franz}(2012)}]{2012_Marchand}
\bibinfo{author}{\bibfnamefont{D.~J.~J.} \bibnamefont{Marchand}}
  \bibnamefont{and} \bibinfo{author}{\bibfnamefont{M.}~\bibnamefont{Franz}},
  \bibinfo{journal}{Phys. Rev. B} \textbf{\bibinfo{volume}{86}},
  \bibinfo{pages}{155146} (\bibinfo{year}{2012}),
  \urlprefix\url{https://link.aps.org/doi/10.1103/PhysRevB.86.155146}.

\bibitem[{\citenamefont{Groth et~al.}(2014)\citenamefont{Groth, Wimmer,
  Akhmerov, and Waintal}}]{2014_Groth}
\bibinfo{author}{\bibfnamefont{C.~W.} \bibnamefont{Groth}},
  \bibinfo{author}{\bibfnamefont{M.}~\bibnamefont{Wimmer}},
  \bibinfo{author}{\bibfnamefont{A.~R.} \bibnamefont{Akhmerov}},
  \bibnamefont{and} \bibinfo{author}{\bibfnamefont{X.}~\bibnamefont{Waintal}},
  \bibinfo{journal}{New Journal of Physics} \textbf{\bibinfo{volume}{16}},
  \bibinfo{pages}{063065} (\bibinfo{year}{2014}),
  \urlprefix\url{https://doi.org/10.1088%2F1367-2630%2F16%2F6%2F063065}.

\bibitem[{\citenamefont{Lin and Hayami}(2016)}]{2016_Lin}
\bibinfo{author}{\bibfnamefont{S.~Z.} \bibnamefont{Lin}} \bibnamefont{and}
  \bibinfo{author}{\bibfnamefont{S.}~\bibnamefont{Hayami}},
  \bibinfo{journal}{\prb} \textbf{\bibinfo{volume}{93}},
  \bibinfo{pages}{064430} (\bibinfo{year}{2016}).

\end{thebibliography}
\end{document}